\documentclass[a4paper,12pt]{article}
\pdfoutput=1 

\usepackage{jcappub} 
\usepackage{multirow}
\usepackage[T1]{fontenc}
\usepackage{graphicx}
\usepackage{epsf}
\usepackage{bm,color}
\usepackage{float}
\usepackage{amsmath}
\usepackage{amsfonts}
\usepackage{amssymb}
\usepackage{epstopdf}
\usepackage{natbib}%
\usepackage{subcaption}
\usepackage{hyperref}
\hypersetup{
	bookmarks=true,         
	unicode=false,          
	pdftoolbar=true,        
	pdfmenubar=true,        
	pdffitwindow=false,     
	pdfstartview={FitH},    
	pdftitle={MHI},    
	pdfauthor={Author},     
	pdfsubject={Subject},   
	pdfcreator={Creator},   
	pdfproducer={Producer}, 
	pdfkeywords={keyword1} {key2} {key3}, 
	pdfnewwindow=true,      
	colorlinks=true,       
	linkcolor=red,          
	citecolor=cyan,        
	filecolor=magenta,      
	urlcolor=blue,           
	linktocpage=true
}
\providecommand{\U}[1]{\protect\rule{.1in}{.1in}}
\newcommand{\be}{\begin{equation}}
	\newcommand{\ee}{\end{equation}}

\newcommand{\mincir}{\raise
	-3.truept\hbox{\rlap{\hbox{$\sim$}}\raise4.truept\hbox{$<$}\ }}
\newcommand{\magcir}{\raise
	-3.truept\hbox{\rlap{\hbox{$\sim$}}\raise4.truept\hbox{$>$}\ }}

\def\bea{\begin{eqnarray}}
	\def\eea{\end{eqnarray}}
\def\ba{\begin{array}}
	\def\ea{\end{array}}

\def\beq{\begin{equation}}
	\def\eeq{\end{equation}}

\newcommand{\eq}{Eq.\eqref}
\newcommand{\fig}{Fig.\ref}
\usepackage{ulem}

\title{Confronting Mukhanov Parametrization of Inflationary Equation-of-State with ACT-DR6}
\author{Barun Kumar Pal}
\emailAdd{terminatorbarun@gmail.com}
\affiliation{Netaji Nagar College For Women, Kolkata-700092, West Bengal, India}
\abstract {We provide a simple yet effective semi-analytical approach to confront Mukhanov Parametrization of inflationary equation-of-state, $1+\omega =\frac{\beta}{({N}+1)^\alpha}$, with the latest ACT-DR6 data employing Hamilton-Jacobi formulation. We find that the equation-of-state formalism renders excellent fit to the most recent observational data. In the process we are also able to put stringent constraint on the two model parameters. In order to constrain the model parameters $\alpha$ and $\beta$ we have also made use of the recent finding $r<0.032$. We have further utilized results from the joint analysis of ACT-DR6, Planck-2018 and DESI-Y1 data to find the observationally viable region for $\alpha$ and $\beta$. We  have also employed the predictions on primordial gravitational  waves from forthcoming CMB missions in the likes of CMB-S4 and LiteBIRD  along with results from the combined analysis of ACT-DR6, Planck-2018 and DESI-Y1 data to further restrict the model parameters. We find that detection of gravitational  waves would narrow the viable parameter space  for Mukhanov parametrization.  But in the absence of detection of primordial tensor mode signal by those CMB missions parameter space is reduced significantly for $\beta$, while the range of $\alpha$ remaining almost constant. In addition  we observe that, $\alpha$ is primarily dependent  on the observationally viable range for scalar spectral index while other model parameter $\beta$ is 	resting heavily on the restriction upon the tensor-to-scalar ratio. Moreover, we find that equation-of-state formalism has a wide range of parameter values consistent with recent observational data along  with  futuristic CMB missions in the likes of CMB-S4 and LiteBIRD. Interestingly, our analysis shows that the joint ACT-DR6, Planck-2018, DESI-Y1, and BICEP/Keck-2018 constraints restore the viability of the monomial class of inflationary models represented by $\alpha\simeq 1$, while ruling out very small-field models, $\alpha\geq 3$. We have also found classical Starobinsy model just lie within $95\%$ confidence level boundary  of the joint analysis of ACT-DR6, Planck-2018, DESI-Y1 and BICEP/Keck-2018 data. 
}
\begin{document}
	\maketitle
	\flushbottom	
	

\section{Introduction}\label{sec1}
Inflation, since it has started its journey more than four decades ago, has surprised us in many ways. The most fascinating among them being production of nearly scale invariant curvature perturbation along with primordial gravitational waves through tensor perturbation. Scale dependence  of scalar curvature perturbation was measured by WMAP \cite{spergel2007} later improved by Planck \cite{akrami2020planck, aghanim2020planck, ade2014planck}. Whereas current bound on the amplitude of primordial gravitational waves is $r<0.032$ \cite{tristram2022improved}, set by BICEP/Keck. Nowadays scalar spectral index along with tensor-to-scalar ratio has come up as eliminators between different viable inflationary models. Recent analyses of the Atacama Cosmology Telescope Data Release 6 (ACT-DR6) \cite{louis2025atacama,  calabrese2025atacama, Naess2025ACTDR6Maps, Atkins2025ACTDR6Covariance} present high-signal temperature, polarization and lensing measurements that are broadly consistent with Planck results while adding crucial high-$\ell$ information that improves constraints on spectral and lensing parameters. The data from Atacama Cosmology Telescope (ACT)  has provided strong indication towards cosmic inflation. It has also come up with a strong indication toward a nearly scale invariant scalar power spectrum, yielding $n_{_S} = 0.9666 \pm 0.0077$, whereas the earlier constraint was $n_{_S} = 0.9649 \pm 0.0042$ 
from Planck-2018 analysis \cite{akrami2020planck,aghanim2020planck}. Furthermore, the latest data release from Planck has pushed this value upward to $n_{_S} = 0.969 \pm 0.0035$ \cite{tristram2024}. This value increases further to $n_{_S} = 0.9709 \pm 0.0038$ when ACT-DR6 is combined with Planck-2018 data. Appending  DESI-Y1 data \cite{adame2025desi, adame2025desi1} on top of the combination of ACT-DR6 and Planck-2018 data provides additional increment for scalar spectral index which is now pushed to $n_{_S}=0.9743\pm0.0034$ \cite{louis2025atacama,  calabrese2025atacama, qu2025unified}. 

The updated estimate of the scalar spectral index favours inflationary models  with de-Sitter like features, which are likely to produce almost scale invariant curvature perturbation. These results put the classical Starobinsky model \cite{starobinsky1980, Lust2024} under significant observational tension. Concurrently, they have motivated a systematic re-examination and refinement of various inflationary frameworks in the light of recent and forthcoming data \cite{KalloshLindeRoest2025,dioguardi2025palatini,dioguardi2025fractional,hjquasi2025,heidarian2025alpha,ferreira2025bao,linde2025alexei,haque2025minimal,okada2025smooth,haque2025act, qiu2025implications,mohammadi2025power, safaei2024observational, pal2025mutated, ribeiro2025cosmic}. Further the futuristic space mission LiteBIRD\cite{litebird2023probing,ghigna2024litebird} and ground based experiment CMB-S4 \cite{abazajian2022cmbS4,belkner2024cmbs4} are anticipating to detect primordial gravitational waves within the range $0.003<r<0.032$. The non-detection of gravitational waves will set the upper bound as $r<0.002$ or $r<0.001$ by LiteBIRD or CMB-S4 respectively. The outcome of combined analysis of ACT-DR6, Planck-2018 and DESI-Y1 data if considered along with the futuristic missions LiteBIRD or CMB-S4 may have the potential to  eliminate numerous inflationary models by substantially narrowing  the observational window. 

In this work, we reassess the observational viability of inflationary scenarios formulated using the Mukhanov equation-of-state  parametrization (EOS, henceforth) \cite{mukhanov2013quantum}. The EOS framework offers a streamlined and model-independent approach that effectively captures a broad spectrum of inflationary dynamics without invoking a specific inflaton potential or modified gravity theory.
Mukhanov’s EOS parametrization was originally introduced as a minimal and general framework capable of reproducing the predictions of a wide class of inflationary potentials that includes power-law/monomial, small-field, hilltop and plateau-type scenarios—via a simple and analytically tractable form of the EOS. Rather than committing to a specific microscopic Lagrangian, this parametrization encodes the inflationary background evolution in terms of a small set of effective parameters. This parametrization offers a model independent and phenomenological framework that connects inflationary dynamics directly to observable quantities such as scalar spectral index, amplitude of primordial gravitational waves, without assuming a specific potential or micro-physical mechanism. This approach is particularly useful when confronting theoretical models with high-precision CMB constraints. While exploring inflationary scenario through defining inflaton potential has the advantage of linking it with the high energy physics in a simple manner. Equation-of-state formalism  is an alternative approach for the investigation of cosmic inflation scenario without entering into the details of underlying physics.  This approach  provides us with a simple but elegant way to encounter the inflationary predictions with observations. Although expressed at the EOS level, the parametrization is not disconnected from fundamental physics. It can be mapped to effective single-field inflation with a reconstructed potential \cite{martin2016observational}. Mukhanov’s original works demonstrated that the parametrized EOS corresponds to potentials that interpolate between well-known inflationary classes (e.g., plateau-type and power-law–like behaviours). Hence, the framework has an underlying scalar-field interpretation that ensures physical consistency. Though purely phenomenological, Mukhanov’s parametrization has been widely used in the analysis of inflationary observables, including studies on attractor behaviour, reconstruction methods and general effective descriptions of inflation \cite{Binetruy:2014zya,Garcia-Bellido:2014wfa,Kinney:2015qma,roest2014universality,garcia2014large,barranco2014model,boubekeur2015phenomenological}.

Nowadays,  equation-of-state based parametrizations have substantially strengthened the motivation for adopting this framework in inflationary studies. The original formulation of the inflationary EOS by Mukhanov \cite{mukhanov2013quantum} and the subsequent identification of universality classes in slow-roll dynamics \cite{roest2014universality} demonstrated that a wide variety of potentials can be captured within a small number of phenomenological parameters. This has been complemented by generalized EOS approaches inspired by modified-gravity and effective-field-theory considerations \cite{Bielefeld:2014nza,martin2014}, which allow the EOS framework to emulate a broad class of gravitational models beyond canonical single-field inflation. Moreover, recent work has provided improved analytic mappings between EoS parameters and observable quantities such as $n_s$ and $r$ \cite{Gong:2015qha}, enabling more direct confrontation of phenomenological EOS models with high-precision CMB data. Collectively, these theoretical developments establish the EoS parametrization as a powerful and flexible tool, offering a unified description of inflationary dynamics across diverse underlying theories.

In this work, we use the Planck-2018 results together with the BICEP/Keck constraint $r < 0.032$ to determine the observationally viable region of the model 
parameters appearing in the EOS formalism. 
 We further constrain the parameter space by incorporating forecasted limits on the  tensor-to-scalar ratio from LiteBIRD and CMB-S4 on top of the Planck-2018 data. In addition, we employ the joint constraints from ACT-DR6, Planck-2018 and DESI-Y1 along with the bound $r < 0.032$ to restrict the parameters even further. We also combine the latest ACT-DR6, Planck-2018 and DESI-Y1 data with the forecasted sensitivities of LiteBIRD and CMB-S4 to significantly narrow the observational window for the Mukhanov parametrization. Moreover, we investigate the consequences of a possible non-detection of primordial gravitational waves by LiteBIRD and CMB-S4 within the EOS framework. By incorporating the latest observational bounds and forecasts,  none of which were included in earlier analyses, we provide the first detailed assessment of the viability of Mukhanov's EOS inflation under present and next-generation CMB constraints. We find that imposing $r < 0.001$, motivated by LiteBIRD and CMB-S4 forecasts, forces the parameter $\beta$ into an extremely narrow range, a result not previously reported. Our analysis further shows that achieving the observed scalar spectral index $n_s$ in the presence of such small tensor amplitudes places strong restrictions on the admissible parameter range of the equation-of-state.

\section{Inflationary Dynamics}
The dynamics of a homogeneous scalar field with the potential $V(\phi)$ and Hubble parameter $H(\phi)$, minimally coupled to gravity is governed by the following first order second degree non-linear differential equation \cite{salopek1990, muslimov1990,liddle1994, kinney1997,lidsey1997, barunquasi,videla2017, barun2018mutated, barunmhi, barunmhip}
\bea
\left[H^{\prime}(\phi)\right]^2 -\frac{3}{2M_P^2}
H(\phi)^2&=&-\frac{1}{2M_P^4}V(\phi)\label{qei:hamilton}\\
\dot{\phi}&=&-2M_P^2 H'(\phi).\label{qei:phidot}
\eea
The above two equations govern the inflationary
dynamics and the corresponding acceleration equation may  be rewritten as 
\beq\label{adot}
\frac{\ddot{a}}{a}=-\left[2M_P^2H^{\prime}(\phi)^2-H(\phi)^2\right]\equiv H^2(\phi)\left[1-\rm\epsilon_{_H}\right],
\eeq
where we have defined $ \epsilon_{_{H}}$ as
\beq\label{epsilon} \epsilon_{_{H}}=2 M_P^2\left(\frac{H'(\phi)}{H(\phi)}
\right)^2.
\eeq 
Accelerated expansion  happens when $\ddot{a}>0$  and this is uniquely determined by the condition $\epsilon_{_{H}}<1$ and it ends exactly at  $\epsilon_{_{H}}=1$. It is noteworthy to mention here  that the requirement leading to a violation of the strong energy condition, $\rho+3P<0$, is uniquely determined by the magnitude of $\epsilon_{_H}$ alone. We can easily derive the following relation \cite{garcia2014large, lidsey1997, kinney1997, kinney2002}
\bea 
\epsilon_{_{H}}&=&\frac{3}{2}(1+P/\rho)\equiv\frac{3}{2}(1+\omega)
\eea 
where $\omega\equiv P/\rho$, is the equation-of-state parameter. It is to be noted that $\epsilon_{_{H}}$ measures the relative contribution of the inflaton's kinetic energy to its total energy \cite{liddle1994, lidsey1997, kinney1997,Copeland1998, kinney2002, garcia2014large, barunquasi2025} which may be seen  from the following relation
\bea
\epsilon_{_{H}}&\equiv&2M_P^2\left(\frac{H'(\phi)}{H(\phi)}\right)^2=\frac{\dot{\phi}^2}{2H^2}.
\eea
Another associated parameter is 
\begin{equation}\label{eta}
\eta_{_{H}}\equiv2M_P^2\frac{H''(\phi)}{H(\phi)}=- \frac{\ddot{\phi}}{H\dot{\phi}} 
\end{equation}
which measures the ratio of the field’s acceleration relative to the friction acting on it due to the expansion of the universe \cite{lidsey1997}. These two parameters are not the usual slow-roll parameters, but slow-roll approximation holds when these parameters are small in comparison to unity.

The amount of inflation is represented by number of e-foldings defined as 
\beq\label{efol}
N(t)\equiv \ln\frac{a(t_{\rm end})}{a(t)}=\int_{t}^{t_{\rm end}} H(t)dt
\eeq 
where $t_{\rm end}$ is the time when inflation comes to an end.  $N$ has been defined  in such a way that at the end of inflation $N=0$ and $N$ increases as we go back in time. The observable parameters are generally evaluated when there are $50-60$ e-foldings still left before the end of inflation. Though total number of e-foldings could be much larger. During this observable period inflationary equation-o-state may be assumed almost constant. The \eq{efol} can be rewritten as a function of the scalar field as follows
\beq\label{nphi}
N(\phi)=-\frac{1}{M_P^2}\int_{\phi}^{\phi_{\rm  end}}\frac{H(\phi)}{2H'(\phi)}\ d\phi=\frac{1}{M_P}\int_{\phi_{\rm  end}}^{\phi}\frac{1}{\sqrt{2\epsilon_{_{H}}}}\ d\phi=\int_{\phi_{\rm  end}}^{\phi}\frac{1}{\epsilon_{_{H}}}\frac{H'(\phi)}{H(\phi)}\ d\phi
\eeq
where $\phi_{\rm end}$ is the value of the scalar field at the end of inflation.

In practice \eq{qei:hamilton} and \eq{qei:phidot} is very complicated due its non-linearity and without a particular Hubble parameter it is very challenging to get the analytical solution.  On the other hand once $H$ has been specified it is possible  to analyze those equations.
\section{Mukhanov Parametrization}\label{mukhanov}
Recently it has been demonstrated that inflationary paradigm can be efficiently dealt by  parametrization the inflationary equation-of-state  as a function of e-foldings without invoking particular form of the inflaton potential or the Hubble parameter.  Primary focus was to develop a model independent framework for the investigation and confrontation of cosmic inflation with recent observations. The parametrization is simply given by \cite{mukhanov2013quantum}
\beq\label{eos}
1+\omega=\frac{\beta}{({N}+1)^\alpha}
\eeq 
where $\alpha \ \mbox{and} \ \beta$ are two dimensionless positive constants and $N$ is the number of e-foldings.  The $\beta$ primarily controls the amplitude of primordial gravitational waves. While $\alpha$ shapes the scalar tilt, and hence $n_{_S}$. This form of the equation-of-state, in principle, can encompass a wide range of inflationary scenario with various observational predictions \cite{gariazzo2017primordial} without entering into the details of underlying high energy physics. Although, initially it was advertised as model independent way-out for cosmic inflation, later proved  that specification of particular equation-of-state indirectly leads to a specific scalar field potential \cite{martin2016observational, barunmeos2025}.  In the next section we shall deduce  the expressions for potential and Hubble parameter associated with the equation-of-state \eq{eos}.

\section{From Equation-of-State to Scalar Field Potential}
In this section we derive associated inflationary Hubble  parameter and potential following Ref.\cite{barunmeos2025}. From the \eq{nphi} we can obtain 
\beq\label{hnphi}
dN=\frac{1}{\epsilon_{_{H}}}\frac{H'(\phi)}{H(\phi)}\ d\phi=\frac{1}{\epsilon_{_{H}}}\frac{dH}{H},
\eeq
which upon integration leads to the following Hubble parameter 
\beq\label{hn}
{H(N)}=H_1\exp\left(\int\epsilon_{_{H}}({N}){dN}\right) 
\eeq
where $H_1$ is the constant of integration. The evolution of scalar field  in terms of e-foldings is found to be
\beq\label{phi-n}
{\phi}=\left\{
\begin{array}{lll}
	{\phi}_{\rm end}+\sqrt{3\beta}{M_P}\ln{(1+N)} & \mbox{if } \alpha =2 \\
{\phi}_{\rm end}+	\frac{\sqrt{3\beta}}{1-\alpha/2}{M_P}\left(\left(1+{N}\right)^{1-\alpha/2} -1\right) & \mbox{if } \alpha\neq2 .
\end{array}
\right.
\eeq

The integration of \eq{hn} gives rise to the Hubble parameter 
\beq\label{hubble-n}
{H}=\left\{
\begin{array}{lll}
	{H}_0\left(1+{N}\right)^{\frac{3\beta}{2}}  & \mbox{if } \alpha =1 \\
	{H}_0\exp\left(\frac{3\beta}{2(1-\alpha)}\left(1+{N}\right)^{1-\alpha}\right)  & \mbox{if } \alpha\neq1 .
\end{array}
\right.
\eeq
Combination of \eq{phi-n} and \eq{hubble-n} leads to the Hubble parameter in terms of the scalar field, 
\beq
{H}=\left\{
\begin{array}{lll}\label{hphi}
	{H}_0\left(2 \sqrt{3 \beta }\right)^{-3 \beta } (\phi/{M_P}) ^{3 \beta }  & \mbox{if } \alpha =1 \\
	{H}_0	\exp \left(-\frac{3\beta}{2}   \exp \left(-\frac{\phi }{\sqrt{3 \beta }{M_P}}\right)\right) & \mbox{if } \alpha=2\\
	{H}_0	\exp \left(-\frac{3 \beta \left(\frac{2-\alpha }{2 \sqrt{3 \beta }}\right)^{\frac{2-2 \alpha }{2-\alpha }}}{2 (1-\alpha )}\ (\phi/{M_P})^\frac{2-2 \alpha }{2-\alpha }\right)  & \mbox{if } \alpha\neq1,2
\end{array}
\right.
\eeq
Finally substituting   \eq{hphi} into \eq{qei:hamilton} we obtain the desired expression for the potential 
\beq\label{vphi}
V(\phi)=\left\{
\begin{array}{lllll}
	{M_P}^4	\ 2^{-6 \beta } 3^{1-3 \beta } \beta ^{-3 \beta } \left(\frac{\phi }{M_P}\right)^{6 \beta -2} \left(\left(\frac{\phi }{M_P}\right)^2-6 \beta ^2\right)& \mbox{if } \alpha =1 \\
	{M_P}^4\ \frac{3}{2} \left(2 e^{\frac{2 \phi\rm /M_P }{\left(\sqrt{3} \sqrt{\beta }\right)}}-\beta \right) \exp \left(-3 \beta  e^{-\frac{\phi \rm / M_P}{\left(\sqrt{3} \sqrt{\beta }\right) }}-\frac{2 \phi \rm /M_P }{\left(\sqrt{3} \sqrt{\beta }\right)}\right)& \mbox{if } \alpha =2 \\
	{M_P}^4\ \exp \left(\frac{2^{-\frac{2 (\alpha -1)}{\alpha -2}} 3^{\frac{1}{2-\alpha }} \beta  \left(-\frac{(\alpha -2) \phi }{\sqrt{\beta } M_P}\right)^{\frac{2 (\alpha -1)}{\alpha -2}}}{\alpha -1}\right)\left(3- \frac{2^{\frac{2-3\alpha}{\alpha -2}} 9^{\frac{1}{2-\alpha }} \beta ^2 \left(-\frac{(\alpha -2) \phi }{\sqrt{\beta } M_P}\right){}^{\frac{4 (\alpha -1)}{\alpha -2}}}{(\alpha -2)^2 \left(\frac{\phi }{\rm M_P}\right)^2}\right) & \mbox{if } \alpha\neq1,2.
\end{array}
\right.
\eeq
Therefore, the equation-of-state \eq{eos} effectively corresponds to the scalar-field potential given in \eq{vphi}. Hence, specifying an equation-of-state indirectly selects a particular inflationary model, as also emphasized in \cite{martin2016observational}.

\section{Inflationary Observables in Equation-of-State Formalism}
With number of e-foldings as new time variable, we can express all the inflationary observables in terms of the equation-of-state \cite{mukhanov2005, garcia2014large},   
\bea
n_{_S} &\simeq&1 -3\left(1+\omega\right) + \dfrac{d}{d{N}}\ln\left(1+\omega\right)\\
\alpha_{_S} &\simeq& 3\dfrac{d}{d{N}}\left(1+\omega\right)-\dfrac{d^2}{d{N}^2}\ln\left(1+\omega\right)\\
r&\simeq&24\left(1+\omega\right)\\
n_{_T}&\simeq&-3\left(1+\omega\right)\\
\alpha_{_T} &\simeq&3\dfrac{d}{d{N}}\left(1+\omega\right)
\eea
where $n_{_T}$,  $\alpha_{_S}$, $\alpha_{_T}$ are tensor spectral index, running of scalar and tensor spectral indices respectively and the results are correct up to the first order in slow-roll parameters. The above observable quantities are evaluated at the time of horizon crossing i.e. when there are 50-60 e-foldings still left before the end of inflation. In this work we have only used scalar spectral index and tensor-to-scalar ratio.  


\section{Confrontation with Observations}
For the confrontation of Mukhanov parametrization of inflationary EOS with observations we have  focused mainly on the scalar spectral index and the tensor-to-scalar ratio using their latest available constraints from various observational data in the likes of Planck-2018, ACT-DR6, DESI-Y1, CMB-S4, LiteBIRD \cite{ade2021improved,tristram2022improved, louis2025atacama,  calabrese2025atacama,adame2025desi1, adame2025desi, belkner2024cmbs4, abazajian2022cmbS4, ghigna2024litebird, litebird2023probing} along with their combinations.  Now for the model under consideration we find that
\bea\label{r_ns}
n_{_S}&\sim& 1-\frac{3\beta}{(1+N)^{\alpha}}-\frac{\alpha}{(1+N)}\\
r&\sim& \frac{24\beta}{(1+N)^{\alpha}}.\label{r_ns1}
\eea
From \eq{r_ns} and \eq{r_ns1} we can get the following relation between tensor-to-scalar ratio and scalar spectral index,
\beq
\frac{r}{8}\sim \left(1-n_{_S}\right)-\frac{\alpha}{(1+N)}\label{rns}.
\eeq
Again \eq{r_ns} can be represented as,
\bea
n_{_S}^{\rm lower}&\lesssim 1-\frac{3\beta}{(1+N)^{\alpha}}-\frac{\alpha}{(1+N)}\lesssim n_{_S}^{\rm upper}\label{nslu},
\eea
where $n_{_S}^{\rm upper/lower} $ correspond to upper/lower bounds of the scalar spectral index from observational point of view. We shall be using the 2-$\sigma$ upper and lower bounds of spectral index from Planck-2018, ACT-DR6 and DESI-Y1 and their combination in order to constraint the model parameters. Also from \eq{r_ns1} we can derive the following inequality, 
\bea
0< \beta&\lesssim& \frac{(1+N)^{\alpha}}{24}\ r^{\rm upper}\label{rlu}, 
\eea
where $r^{\rm upper}$ is the current upper bound of the amplitude of primordial gravitational waves. In the following we shall use \eq{nslu} and \eq{rlu} in order to put stringent constraint on the model parameters $\alpha$ and $\beta$ using different results from various observations and their combinations.  
\subsection{Confrontation with Planck-2018 Results}
\begin{figure}
	\centering
	\begin{subfigure}{.5\textwidth}
		\centering
		\includegraphics[width=.97\linewidth,height=7cm]{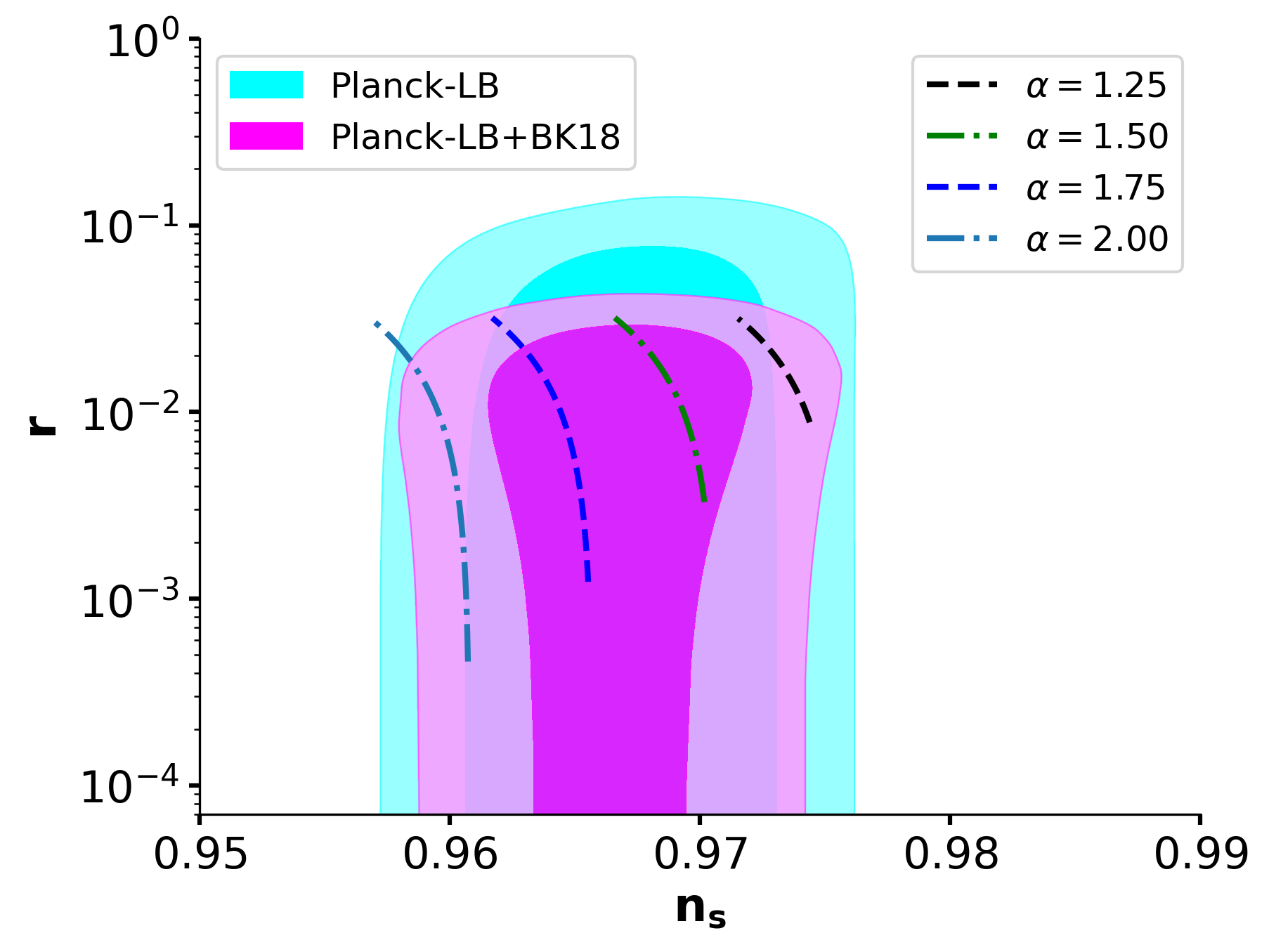}
		\caption{N=50}
	\end{subfigure}%
	\begin{subfigure}{.5\textwidth}
		\centering
		\includegraphics[width=.97\linewidth,height=7cm]{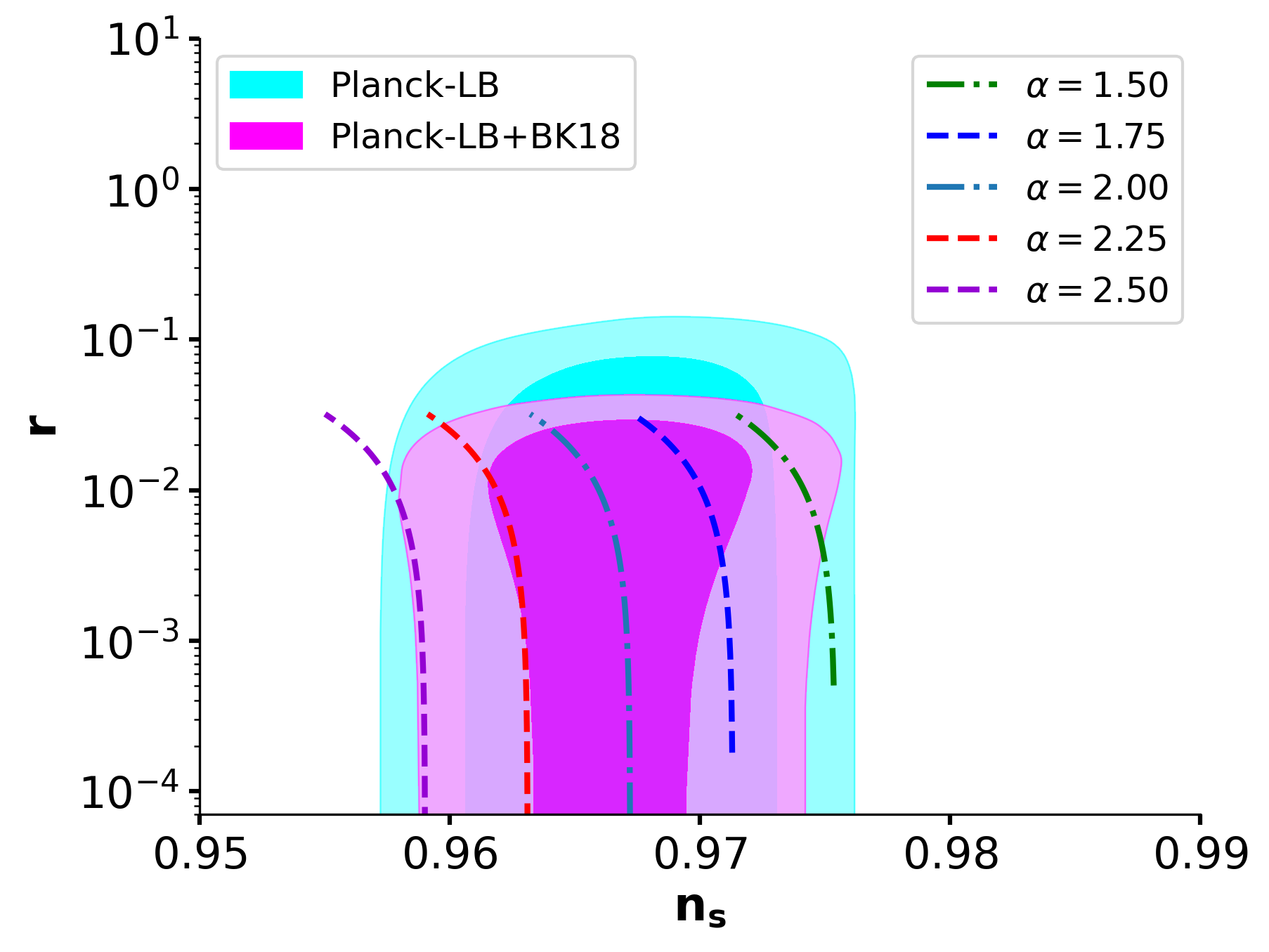}
		\caption{N=60}
	\end{subfigure}
	\caption{Constraints on the scalar spectral index $n_s$ and the tensor-to-scalar ratio $r$ from Planck 2018. The cyan and magenta shaded regions denote the $68\%$ and $95\%$ confidence contours from {Planck+Lensing+BAO} and {Planck+Lensing+BAO+BK18}, respectively. The dashed curves correspond to predictions of the EOS inflation model for different values of the parameter $\alpha$ and fixed $N=50, \ 60$. For each fixed value of $\alpha$, there exists a corresponding range of $\beta$ values for which the observational predictions of the EOS model remain consistent with the Planck 2018 data.}
	\label{fig_rns-planck18}
\end{figure}

To start with, we make use of the Planck-2018 results\cite{akrami2020planck, aghanim2020planck} along with the constraint on the amplitude of primordial tensor perturbations, $r<0.032$ \cite{tristram2022improved}, to restrict the model parameters, $\alpha$ and $\beta$. The constraint on scalar spectral index set by Planck-2018 is $n_{_S}=0.9649\pm 0.0042$ \cite{akrami2020planck, aghanim2020planck} and using 2-$\sigma$ upper and lower bounds of scalar spectral index along with $r<0.032$, we can find corresponding  upper and lower bounds of $\alpha$. In \fig{fig_rns-planck18} we have shown predictions from EOS for different values of $\alpha$ at fixed $N=50$ and 60 on top of the $68\%$ and $95\%$ confidence contours from {Planck+Lensing+BAO} (Planck-LB) and {Planck+Lensing+BAO+BK18} (Planck-LB+BK18) data, respectively in the $r-{n_{_S}}$ plane, while varying the other model parameter $\beta$ within its allowed range.  These contours summarize the current observational bounds on the primordial scalar tilt and the amplitude of primordial gravitational waves. The dashed and dot-dashed curves show the predictions of EOS inflation evaluated for a particular value of $\alpha$ and different values of $\beta$. The upper-bound of model parameter $\beta$ has been fixed from the constraint $r<0.032$ \cite{tristram2022improved} while lower limit has been set as $\beta>0$.  From the figure, we observe that a sizeable region of parameter space set by the choice of  number of e-foldings allows the inflationary predictions of the Mukhanov parametrization to remain fully consistent with the Planck-2018 constraints. Comparing between  with the cases $N=50$ and  $N=60$ we observe that later provides a noticeably broader allowed parameter range for the EOS model. This widening occurs because increasing the number of e-foldings shifts the theoretical predictions of $n_{_S}$ and $r$
toward regions more compatible with the observational contours. In particular, a larger $N$ typically brings the scalar spectral index within its present observational  value and suppresses the tensor-to-scalar ratio, thereby accommodating a wider range of the model parameter $\beta$ as well as $\alpha$. As a result, the EOS model enjoys greater flexibility and more viable parameter space at $N=60$ compared to $N=50$. 
\begin{figure}
	\centerline{\includegraphics[width=15.cm, height=8cm]{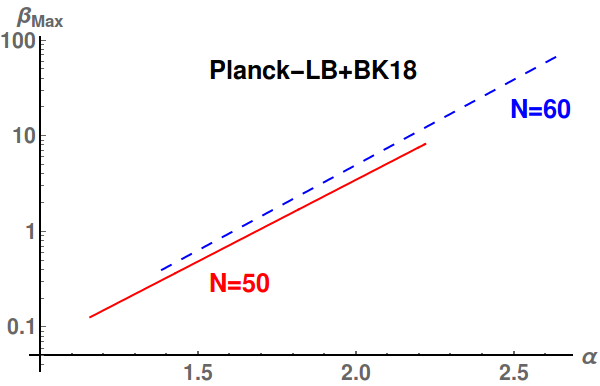}}
	\caption{\label{fig_betamax_planck}Variation of $\beta_{\rm Max}$ with the model parameter $\alpha$ for $N=50$ (solid red line) and $N=60$ (dashed blue curve). For the plot we have considered the constraint on scalar spectral index (2-$\sigma$ upper/lower bound) from Planck-2018 result along with  $r<0.032$.}
\end{figure}
Using 2-$\sigma$ upper and lower bounds of scalar spectral index along with $r<0.032$, we can find corresponding  upper and lower bounds of $\alpha$. We can also  set the upper limit on the other model parameter $\beta$, which has been illustrated in \fig{fig_betamax_planck}. From the figure we see that there is a decent parameter space depending on the number of e-foldings where inflationary predictions from  Mukhanov parametrization are in accord with Planck-2018 result. 
However, inspection of the above plots shows that values of  $\alpha\sim 1$,
which correspond to power-law or monomial inflation models, are fully excluded by the allowed parameter region. In contrast, models characterized by $\alpha \sim 2$, encompassing the classical Starobinsky scenario \cite{starobinsky1980} and the various $\alpha$-attractors \cite{kallosh2013, Kallosh:2013hoa, Kallosh:2013maa, Kallosh:2014rga}, are fully consistent with and strongly favoured by the Planck-2018 observational constraints. Hilltop or small-field inflation in the EOS family corresponds to large $\alpha$ and small $\beta$, which lead to negligibly small tensor amplitudes, remains viable, but in the regime of large $\beta$.
 For $N=60$ and $\alpha\simeq 2.5$, the EOS framework yields predictions that remain within the observationally allowed region. Nevertheless, consistency in this part of parameter space demands comparatively large values of $\beta$, which compensate for the stronger suppression of $r$ induced by the larger $\alpha$.

\begin{figure}
	\centering
	\begin{subfigure}{.5\textwidth}
		\centering
		\includegraphics[width=.97\linewidth,height=7cm]{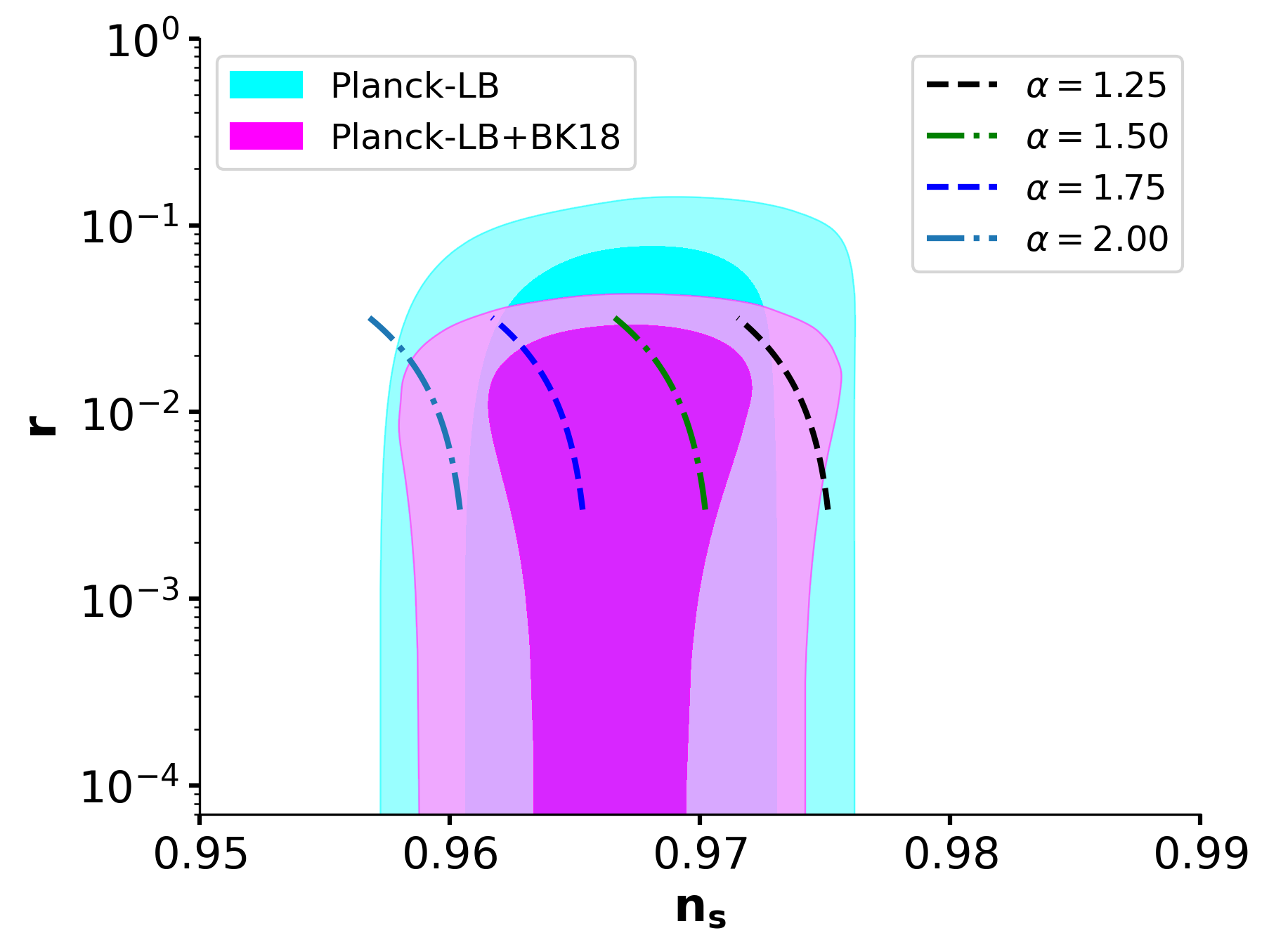}
		\caption{N=50: LiteBIRD Detection}
	\end{subfigure}%
	\begin{subfigure}{.5\textwidth}
		\centering
		\includegraphics[width=.97\linewidth,height=7cm]{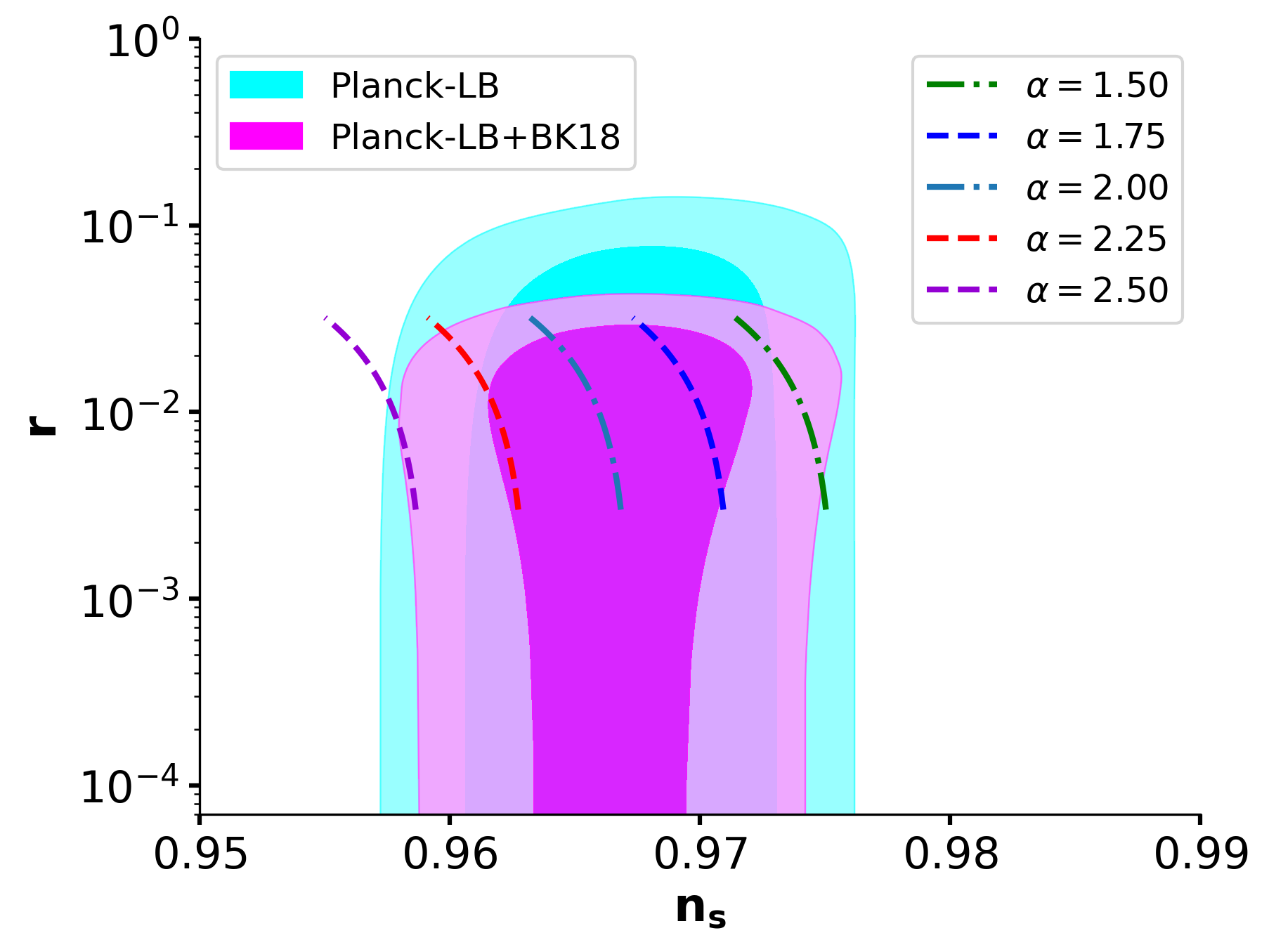}
		\caption{N=60: LiteBIRD Detection}
	\end{subfigure}
	\caption{Constraints on the scalar spectral index $n_s$ and the tensor-to-scalar ratio $r$ from Planck-2018 analysis in combination with Lensing, BAO and BK18. The cyan and magenta shaded regions denote the $68\%$ and $95\%$ confidence contours from {Planck+Lensing+BAO} and {Planck+Lensing+BAO+BK18}, respectively. The dashed curves correspond to predictions of the EOS inflation model for different fixed values of the parameter $\alpha$ and fixed $N=50, \ 60$, while varying $\beta$ within its allowed range $0.003\leq0.032$ as predicted by LiteBIRD. For each fixed value of $\alpha$, there exists a corresponding range of $\beta$ values for which the observational predictions of the EOS model remain consistent with the Planck 2018 data and LiteBIRD sensitivity for a detection of primordial gravitational waves.}
	\label{fig_rns-planck18lb}
\end{figure}

\begin{figure}
	\centering
	\begin{subfigure}{.5\textwidth}
		\centering
		\includegraphics[width=.95\linewidth,height=6cm]{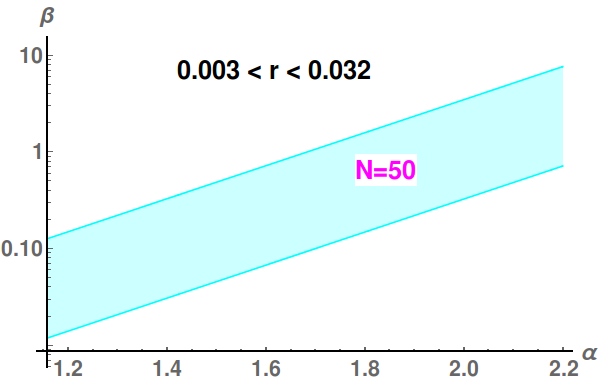}
		\caption{N=50: LiteBIRD detection}
	\end{subfigure}%
	\begin{subfigure}{.5\textwidth}
		\centering
		\includegraphics[width=.95\linewidth,height=6cm]{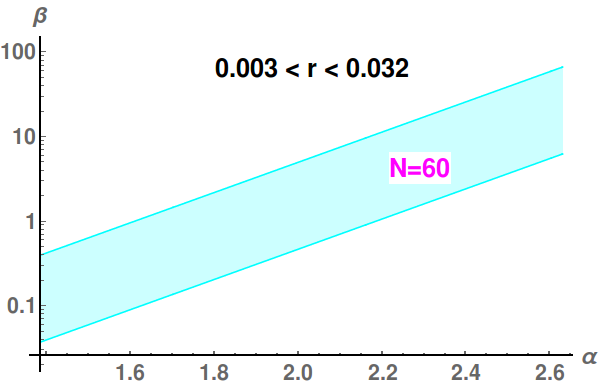}
		\caption{N=60: LiteBIRD detection}
	\end{subfigure}
	\caption{Variation of  $\beta$ with the model parameter $\alpha$ for $N=50$ and $N=60$. For the plot we have considered the constraint on primordial gravity waves   $0.003<r<0.032$ from the sensitivity of futuristic space mission LiteBIRD and along with the recent bound from Planck+Lensing+BAO+BK18 results. Shaded region corresponds to the observationally viable range for $\beta$.}
	\label{fig_betamax_planck_LBD}
\end{figure}

The future CMB space mission LiteBIRD is promising to detect gravitational waves within the range $0.003<r<0.032$ \cite{litebird2023probing,ghigna2024litebird}. In \fig{fig_rns-planck18lb} we have illustrated tensor-to-scalar ratio versus spectral index in the $r-n_{_S}$ plane, for two different values of e-foldings, $N = 50,\ 60$. The dashed and dot-dashed curves represent the predictions of the EOS model for several fixed values of $\alpha$, while the other model parameter $\beta$ has been varied. The allowed range of $\beta$ is determined by imposing the LiteBIRD sensitivity window on the tensor-to-scalar ratio, $0.003 < r < 0.032$, ensuring that the corresponding predictions fall within the detectable range of the experiment. Shaded contour, as before, represents  marginalized $68\%$ and $95\%$ confidence regions in the plane of $r-n_{_S}$ from the Planck+Lensing+BAO results, both with and without the inclusion of BK18 \cite{ade2018constraints, ade2021improved}.  In this case we can set both the limits, upper and lower, on the parameter $\beta$ as depicted in \fig{fig_betamax_planck_LBD}. The shaded area corresponds to the observationally viable region where model parameters are in tune with the combined analysis of Planck+Lensing+BAO+BK18 and  predictions from futuristic space mission LiteBIRD. 
From \fig{fig_rns-planck18lb} and \fig{fig_betamax_planck_LBD}, we see that the observationally viable range of $\beta$ is reduced compared to the case where LiteBIRD sensitivity requirement $r > 0.003$ is not imposed. This reduction occurs because enforcing the lower bound on $r$ excludes those regions of parameter space that would otherwise predict tensor amplitudes below the LiteBIRD detection threshold. Consequently, only those $\beta$ values that simultaneously satisfy the Planck constraints on $r$ -- $n_s$ and produce a tensor-to-scalar ratio within the LiteBIRD sensitivity window remain viable. 
It is evident from the parameter space that power-law/monomial inflation models are fully ruled out, as their predictions lie completely outside the observationally allowed region. But, Starobinsky model once again lie within the observational window. To incorporate hilltop or small-field inflationary models within the EOS formulation in a manner consistent with the Planck-2018 results, while also anticipating a possible detection of primordial gravitational waves by LiteBIRD, one requires a larger number of e-foldings and higher $\beta$ values.

\begin{figure}
		\centering
		\begin{subfigure}{.5\textwidth}
				\centering
				\includegraphics[width=.97\linewidth,height=7cm]{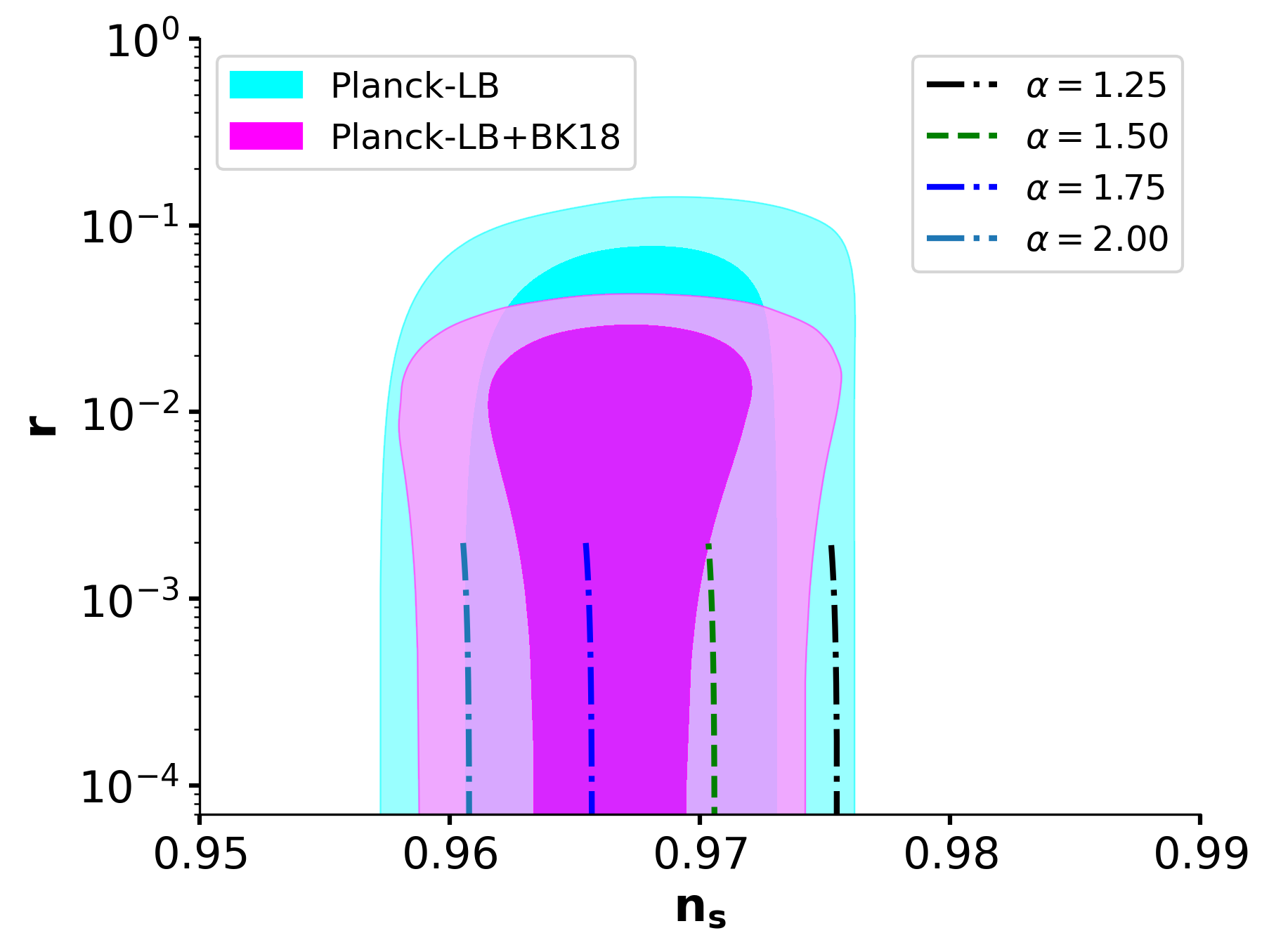}
				\caption{N=50: LiteBIRD Non-Detection}
			\end{subfigure}%
		\begin{subfigure}{.5\textwidth}
				\centering
				\includegraphics[width=.97\linewidth,height=7cm]{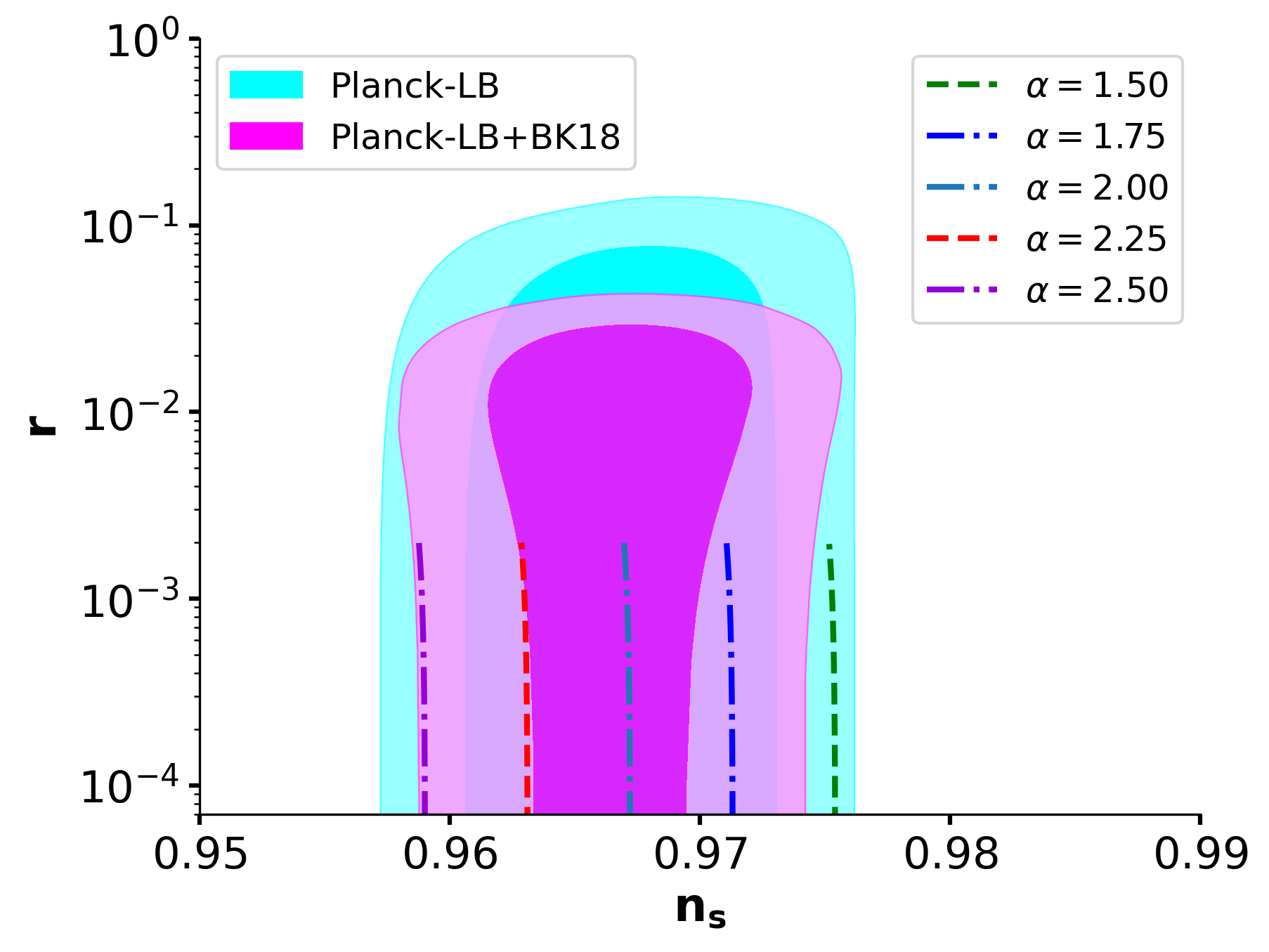}
				\caption{N=60: LiteBIRD Non-Detection }
			\end{subfigure}
		\caption{The cyan and magenta shaded regions  as before denote the $68\%$ and $95\%$ confidence contours from {Planck+Lensing+BAO} and {Planck+Lensing+BAO+BK18}, respectively. The dashed curves correspond to predictions of the EOS inflation model for different values of the parameter $\alpha$ and fixed $N=50, \ 60$ (Left and Right Panel respectively).  The Lower limit of $\beta$ has been set from the LiteBIRD sensitivity for non-detection of primordial gravitational waves $r<0.002$.}
		\label{fig_planck18nlb}
	\end{figure}

\begin{figure}
	\centerline{\includegraphics[width=15.cm, height=8cm]{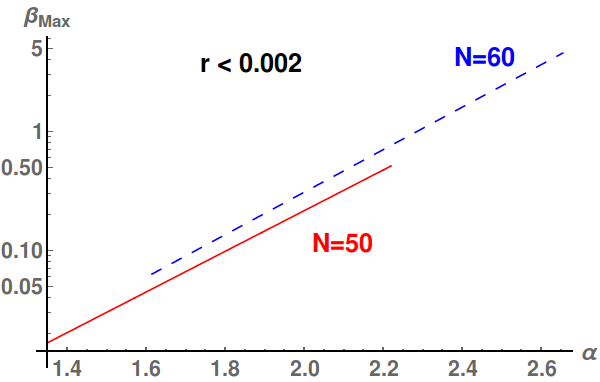}}
	\caption{\label{fig_betamax_plancknlb}Variation of $\beta_{\rm Max}$ with the model parameter $\alpha$ for $N=50$ (solid red line) and $N=60$ (dashed blue curve). For the plot we have considered the constraint on scalar spectral index from Planck-2018 result along with  $r<0.002$ as predicted by LiteBIRD for non-detection of primordial gravitational waves.}
\end{figure}

 However,  if LiteBIRD does not succeed in detecting primordial gravitational waves then $r$ will be pushed back to $r<0.002$ and  we do have new upper-limit of $\beta$ from \eq{rlu}. In \fig{fig_planck18nlb} we have shown EOS prediction in $r-n_{_S}$ plane for two values of number of e-foldings and for different $\alpha$ values, while other model parameter $\beta$ has been varied within its allowed range. We notice that even if LiteBIRD fails to detect primordial gravitational waves, the EOS inflationary framework can still accommodate a broad region of parameter space that remains fully consistent with current observational constraints. But now the parameter space, in particular $\beta$, has been noticeably reduced as we have illustrated  in \fig{fig_betamax_plancknlb}. We again find that the Starobinsky  and the $\alpha$-attractor class (with $\alpha \simeq 2$) naturally fall within the preferred region of the $(\alpha,\beta)$ plane. In comparison, models characterized by effective $\alpha \simeq 2.5\text{--}2.6$ lie close to the observational boundary and can be reconciled with the data only if one assumes both an increased number of e-foldings and relatively large $\beta$ values.

\begin{figure}
 	\centering
 	\begin{subfigure}{.5\textwidth}
 		\centering
 		\includegraphics[width=.97\linewidth,height=7cm]{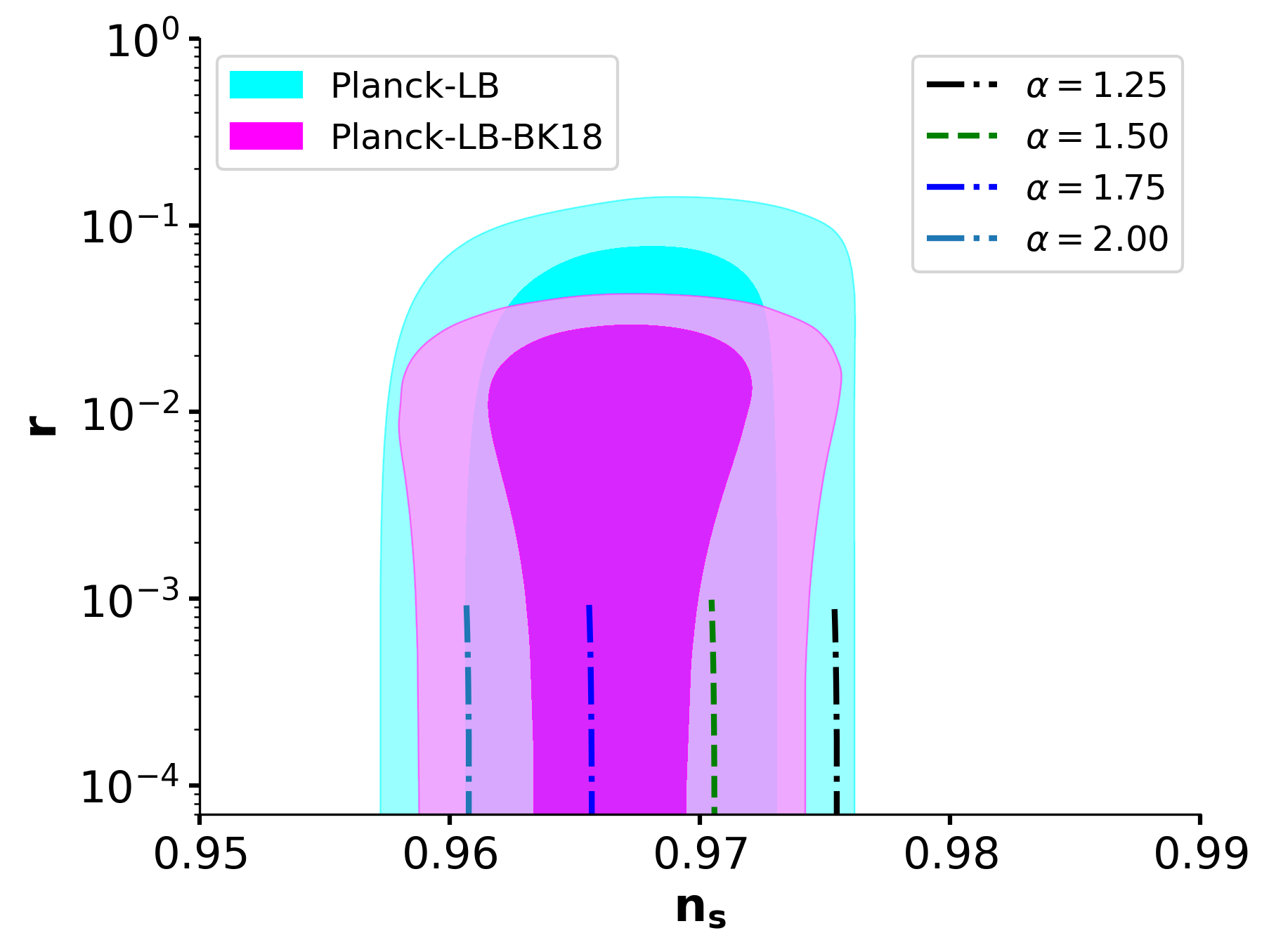}
 		\caption{N=50: CMB-S4 Non-Detection}
 	\end{subfigure}%
 	\begin{subfigure}{.5\textwidth}
 		\centering
 		\includegraphics[width=.97\linewidth,height=7cm]{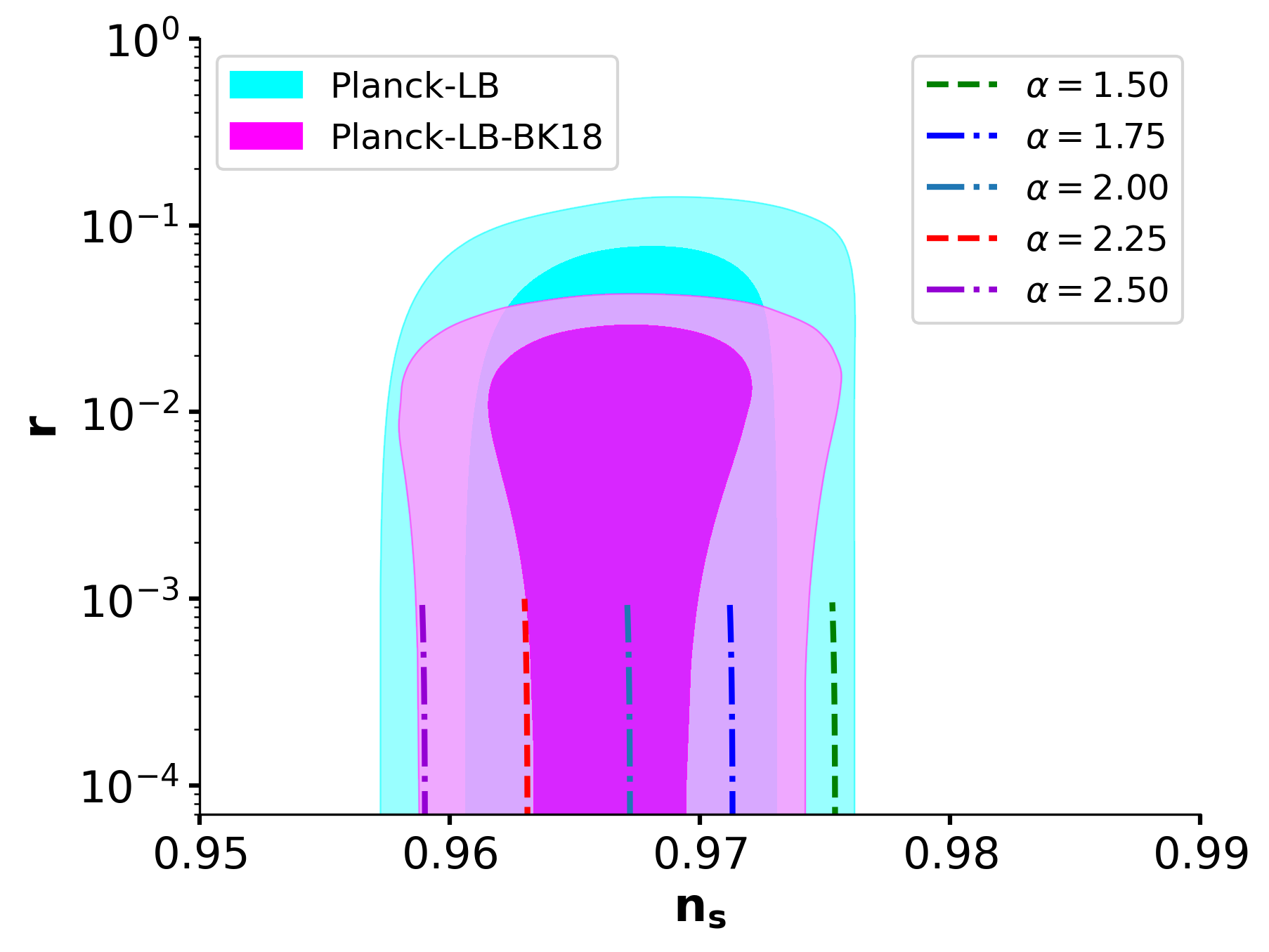}
 		\caption{N=60: CMB-S4 Non-Detection}
 	\end{subfigure}
 	\caption{The $68\%$ and $95\%$ confidence contours from {Planck+Lensing+BAO} and {Planck+Lensing+BAO+BK18} in the $r-n_{_S}$ plane. The dashed curves correspond to predictions of the EOS inflation model for different values of the parameter $\alpha$ and fixed $N=50, \ 60$ (Left and Right Panel respectively).  The Lower limit of $\beta$ has been set from the CMB-S4 sensitivity for non-detection of primordial gravitational waves $r<0.001$.}
 	\label{fig_planck18ns4}
 \end{figure}

\begin{figure}
	\centerline{\includegraphics[width=15.cm, height=8cm]{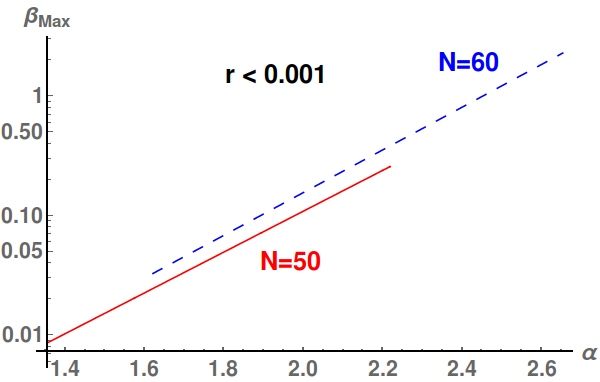}}
	\caption{\label{fig_betamax_planckns4}Variation of $\beta_{\rm Max}$ with the model parameter $\alpha$ for $N=50$ (solid red line) and $N=60$ (dashed blue curve). For the plot we have considered the constraint on scalar spectral index from Planck-2018 result along with  $r<0.001$.}
\end{figure}

 In case of CMB-S4,  a detection of primordial gravitational waves will imply $0.003<r<0.032$, consequently the parameter space will be exactly same as that of  LiteBIRD, and we do not repeat the analysis. However, a non-detection of primordial tensor perturbation will push the limit to $r<0.001$. In that case the parameter space is notably cut down as shown in \fig{fig_planck18ns4} and \fig{fig_betamax_planckns4}. But here also we can get sufficient region for the model parameters, where the EOS predictions will be in accord with the CMS-S4 non-detection of primordial gravitational waves. 

Therefore, from the above analysis, we find that EOS inflation admits a broad region of parameter space in which its inflationary predictions remain in excellent agreement with the Planck-2018 results. Moreover, the model also shows strong compatibility with forecasts from upcoming CMB experiments such as LiteBIRD and CMB-S4. Importantly, we also find that a non-detection of primordial gravitational waves by these future missions will greatly help to constrain the EOS parameter space. Overall, monomial potentials (corresponding to $\alpha \lesssim 1$) are strongly disfavoured by the Planck-2018 data. In contrast, the intermediate range $1.25 \lesssim \alpha \lesssim 2.5$, which corresponds to inflationary models interpolating between power-law and plateau-type scenarios such as Starobinsky or the $\alpha$-attractors, remains fully consistent with current observations. This region also includes models that lie between true plateau inflation (e.g., Starobinsky, $\alpha = 2$) and mildly small-field or shallow hilltop potentials (approaching $\alpha \simeq 2.5$). Importantly, this entire $\alpha$-interval continues to provide an excellent fit when confronted with both Planck-2018 constraints and the forecasted sensitivities of future CMB missions such as LiteBIRD and CMB-S4.


\subsection{Confrontation with 	ACT-DR6 Result} 
\begin{figure}
	\centering
	\begin{subfigure}{.5\textwidth}
		\centering
		\includegraphics[width=.97\linewidth, height=6cm]{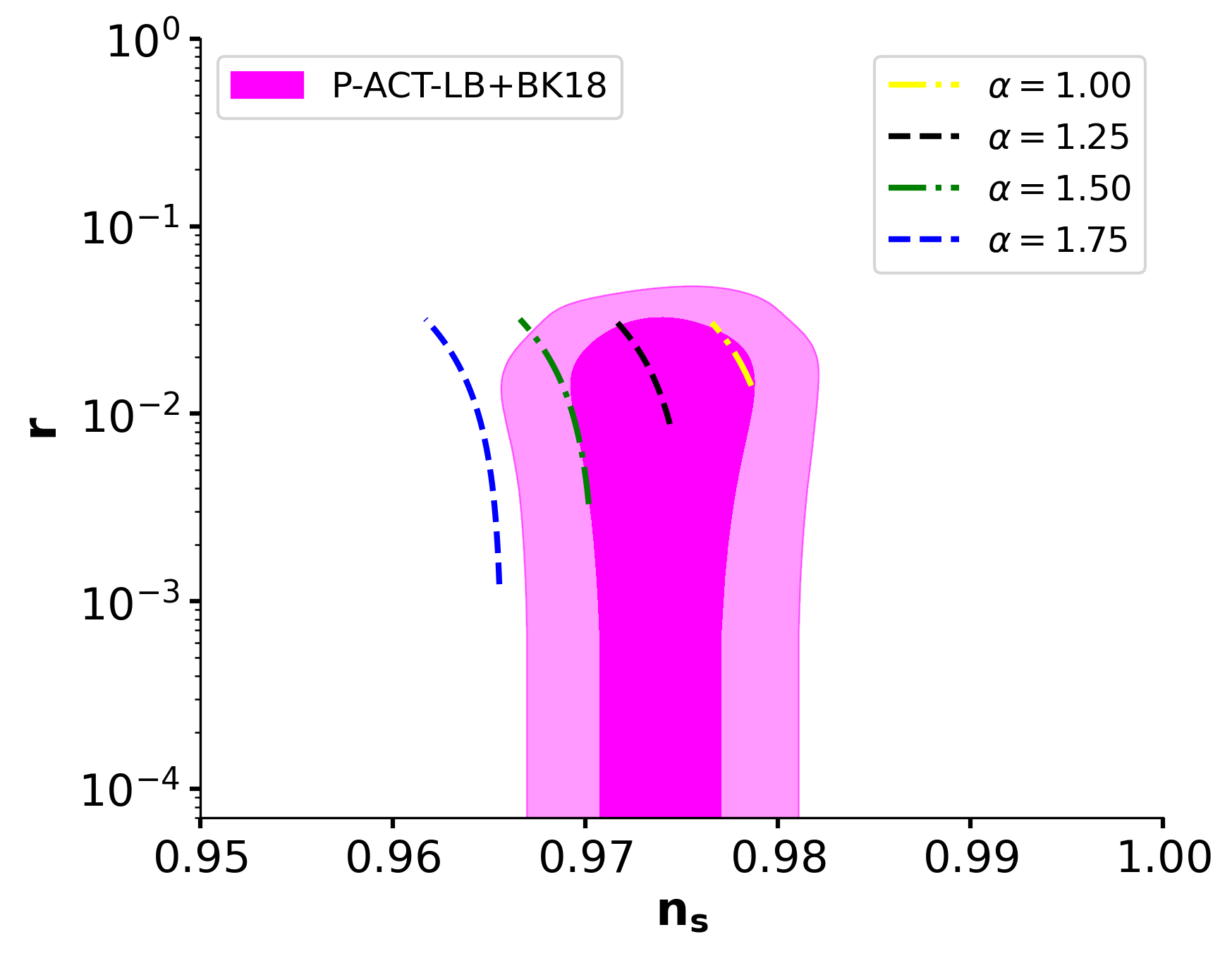}
		\caption{N=50}
	\end{subfigure}%
	\begin{subfigure}{.5\textwidth}
		\centering
		\includegraphics[width=.97\linewidth, height=6cm]{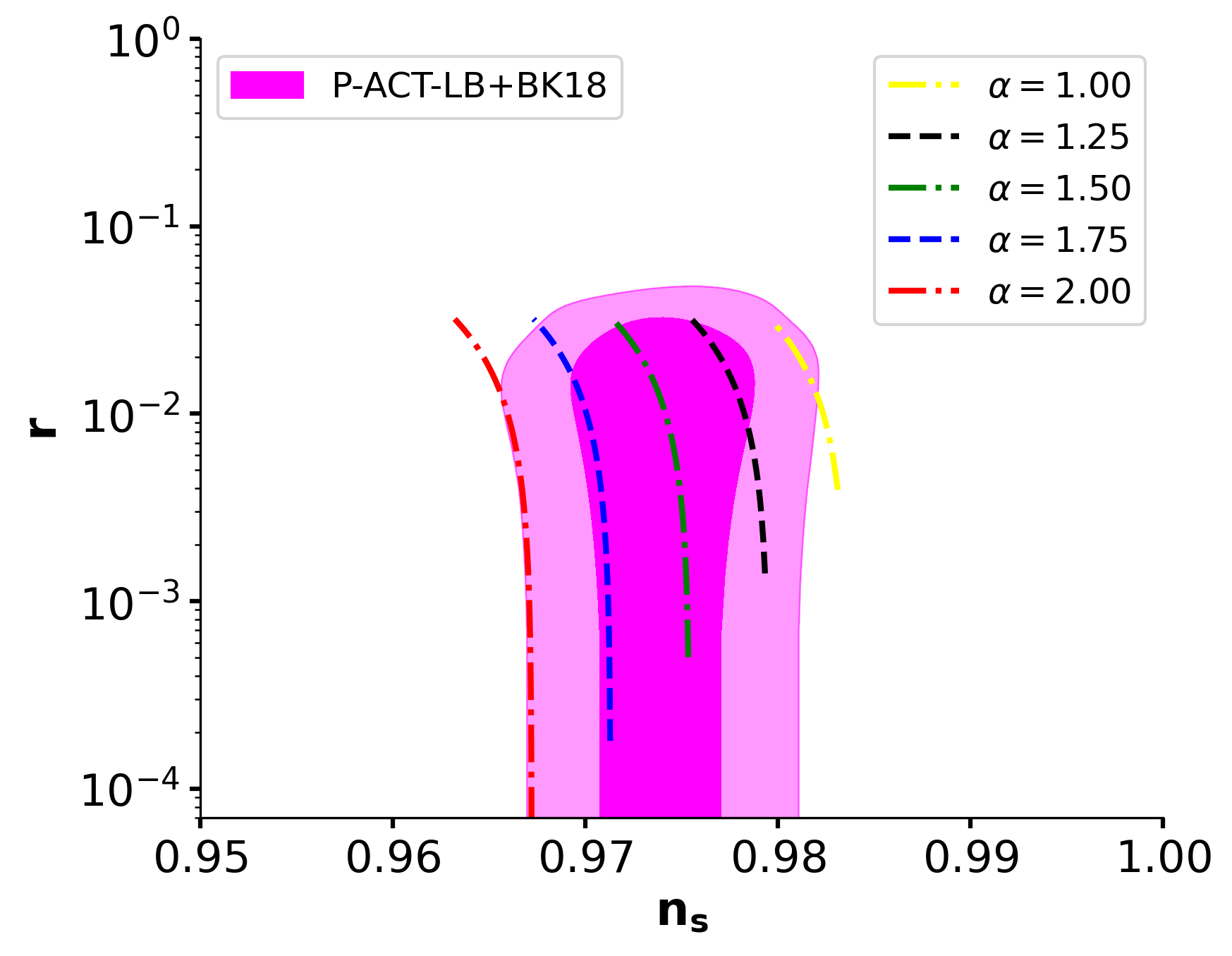}
		\caption{N=60}
	\end{subfigure}
	\caption{Variation of the tensor-to-scalar ratio, $r$, with the scalar spectral index, $n_{_S}$, for two different values of e-foldings, $N = 50,\ 60$. Dashed and dot-dashed lines correspond to predictions from the EOS model for varying values of the model parameter $\beta$ within its allowed range, and fixed $\alpha$ values. The marginalized $68\%$ and $95\%$ confidence regions in the plane of $r-n_{_S}$ from the combined analysis of  ACT, Planck-2018 joint with BK18    and DESI-Y1 data\cite{calabrese2025atacama}. Where the constraint on $r$ is obtained from BK18 data \cite{ade2018constraints} and that of $n_{_S}$ is driven by the combination of Planck, ACT DR6, and DESI-Y1 data.}
	\label{fig_rns-pact}
\end{figure}

\begin{figure}
	\centerline{\includegraphics[width=15.cm, height=8cm]{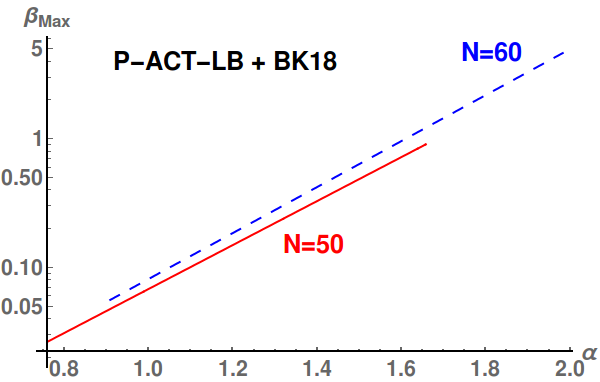}}
	\caption{\label{fig_betamax_pact}Variation of $\beta_{\rm Max}$ with the model parameter $\alpha$ for $N=50$ (solid red line) and $N=60$ (dashed blue curve). For the plot we have considered the constraint on scalar spectral index from the combined analysis of P-ACT-LB data along with  $r<0.032$ from BK18 data.}
\end{figure}

We now move onto the case where ACT-DR6 data is combined with Planck-2018 and DESI-Y1 data. The joint analysis shows ample shift in the scalar spectral index towards scale invariance, $n_{_S}=0.9743\pm0.0034$. This is a very interesting result, as it indicates that the curvature perturbation is even closer to scale invariance than what was inferred from the Planck data. In \fig{fig_rns-pact}, we illustrate that EOS inflation delivers an excellent fit to the observational constraints derived from the combined analysis of Planck, BK18, ACT-DR6 and DESI Y1 data sets. The marginalized joint $68\%$ and $95\%$ confidence contours in $r-n_{_S}$ plane is from the combined analysis of Planck, ACT-DR6 and DESI Y1 data (P-ACT-LB) along with the present constraint on primordial gravitational waves, $r<0.032$, from BK18 \cite{tristram2022improved}. Here the upper limit of $\beta$ has been set from present constraint on tensor-to-scalar ratio. As evident from the \fig{fig_betamax_pact}, the combined analysis further tightens the constraints and substantially shrinks the viable parameter space for EOS. From \fig{fig_rns-pact} and \fig{fig_betamax_pact}, we observe that EOS inflationary models with $\alpha \lesssim 1$ corresponding to power-law/monomial potentials long held to be disfavoured by Planck-2018, now seem to become viable once again for small values of $\beta$. In contrast, accommodating the well-known Starobinsky model ($\alpha\sim 2$) requires a larger number of e-folds, $N \gtrsim 60$ and large $\beta$ values. The joint constraint also indicates that hilltop or small-field inflation models (with $\alpha > 2$) are strongly disfavoured by the data. 

Now further if we use restriction on primordial gravitational waves from the sensitivity of upcoming space mission LiteBIRD, $0.003<r<0.032$, along with the result from combined analysis of P-ACT-LB+BK18, the parameter space remains almost same as that with the constraint $r<0.032$. However, with the updated constraints, the viable interval for the parameter $\beta$ shrinks to a much narrower strip.
\begin{figure}
	\centering
	\begin{subfigure}{.5\textwidth}
		\centering
		\includegraphics[width=.97\linewidth, height=7cm]{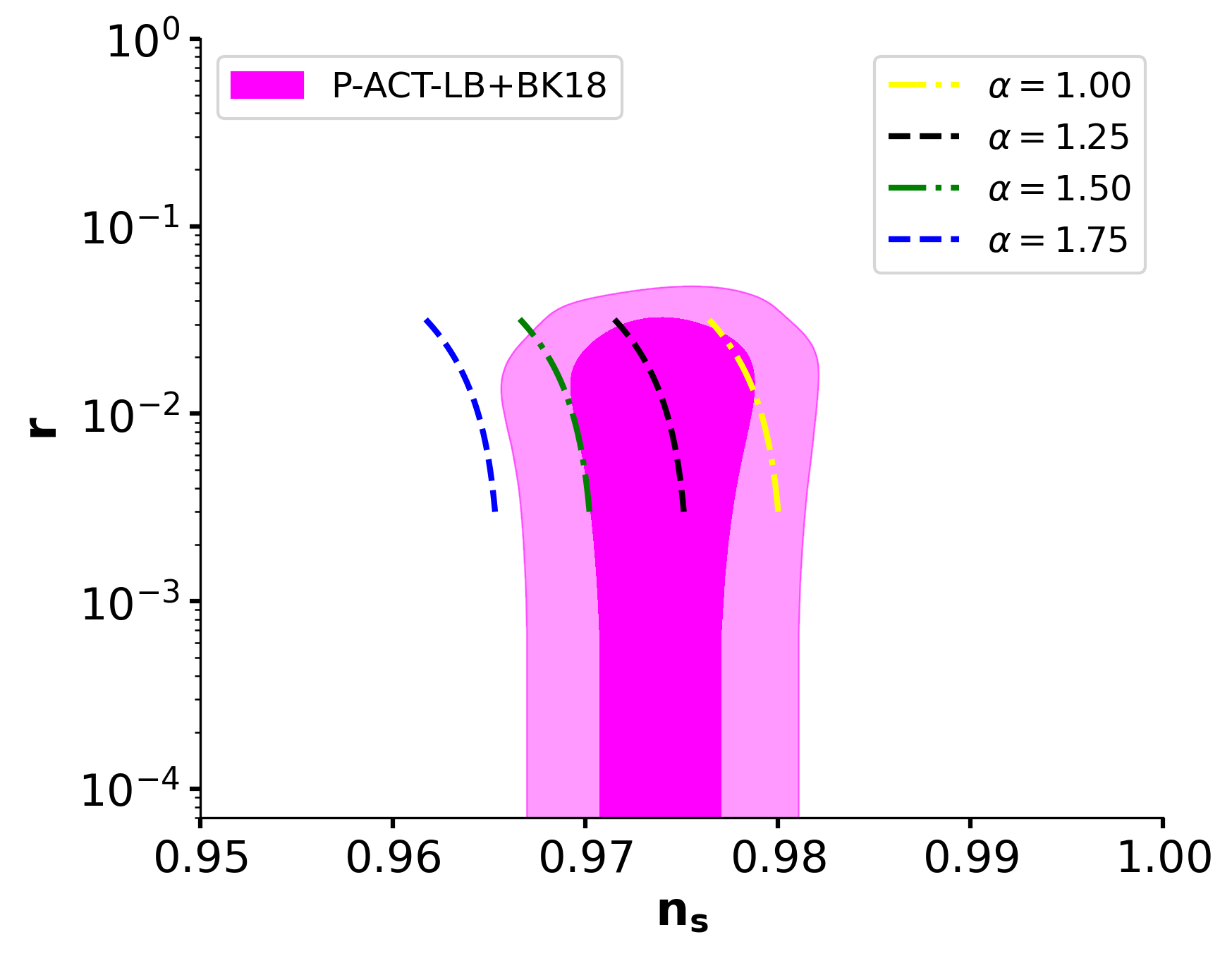}
		\caption{N=50: LiteBIRD Detection}
	\end{subfigure}%
	\begin{subfigure}{.5\textwidth}
		\centering
		\includegraphics[width=.97\linewidth, height=7cm]{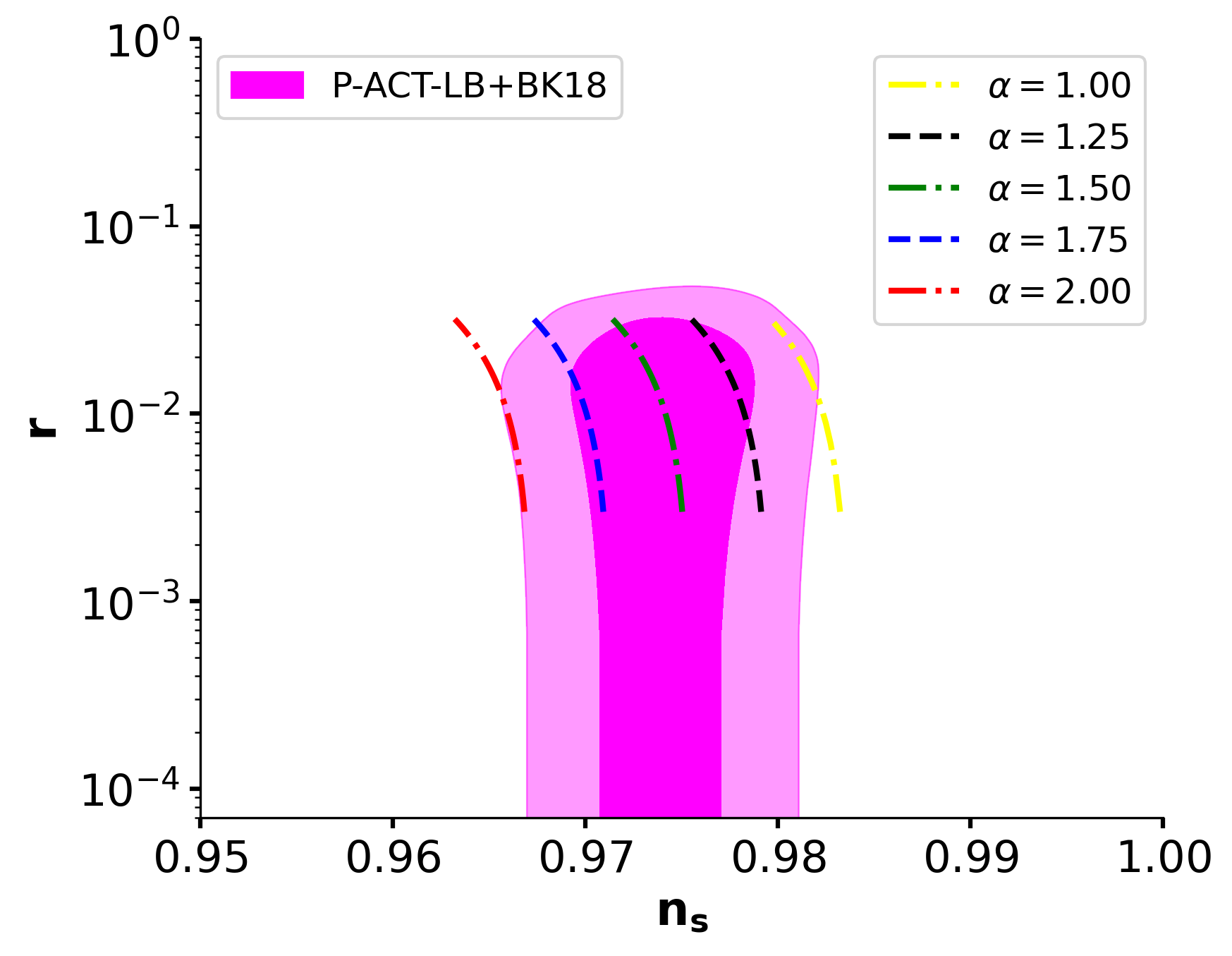}
		\caption{N=60: LiteBIRD Detection}
	\end{subfigure}
	\caption{The $68\%$ and $95\%$ confidence regions in the $r$--$n_{_S}$ plane obtained from the joint analyses Planck+LB+BK18, ACT+LB+BK18, and P-ACT-LB+BK18. The dashed curves denote the EOS inflation predictions for different fixed values of the model parameter $\alpha$, two values of number of e-folding and varying the other model parameter $\beta$ within its allowed range determined by $0.003<r<0.032$. As evident from the figure, the inclusion of ACT data alongside Planck, LiteBIRD, and BK18 substantially reduces the allowed parameter space for EOS inflation.}
	\label{fig_rns-pactlb}
\end{figure}
\begin{figure}
	\centering
	\begin{subfigure}{.5\textwidth}
		\centering
		\includegraphics[width=.95\linewidth,height=6cm]{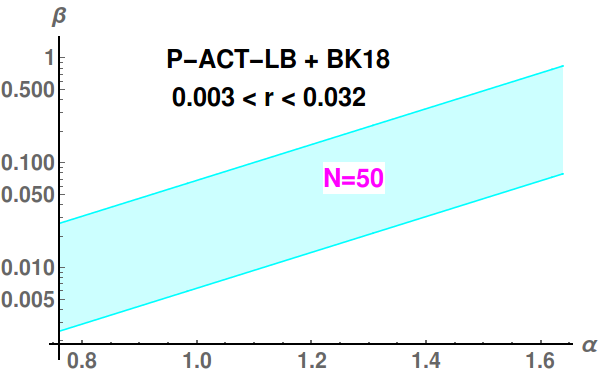}
		\caption{N=50}
	\end{subfigure}%
	\begin{subfigure}{.5\textwidth}
		\centering
		\includegraphics[width=.95\linewidth,height=6cm]{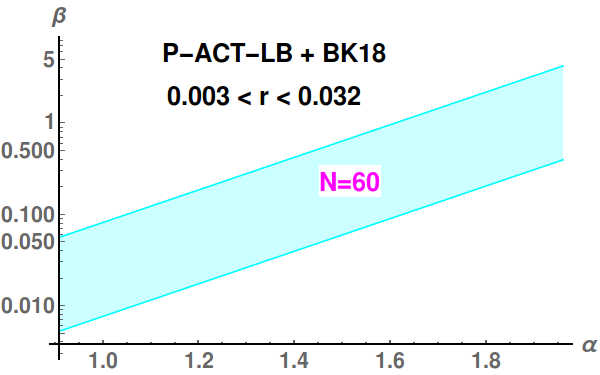}
		\caption{N=60}
	\end{subfigure}
	\caption{Variation of  $\beta$ with the model parameter $\alpha$ for $N=50$ and $N=60$. For the plot we have considered the constraint on primordial gravity waves   $0.003<r<0.032$ from the sensitivity of futuristic space mission LiteBIRD and along with the 2-$\sigma$ upper and lower bounds on scalar spectral index from P-ACT-LB+BK18 results.}
	\label{fig_betamax_pact_LBD}
\end{figure}
\fig{fig_rns-pactlb} illustrates that incorporating ACT data alongside Planck, BK18 and  LiteBIRD sensitivity results into a substantial reduction in the allowed parameter space of EOS inflation, as $\beta$ is now bounded from both above and below. Despite this tightening, the model continues to produce predictions that lie comfortably within the observationally permitted region for various values of $\alpha$ and a range of $\beta$. In \fig{fig_betamax_pact_LBD} we have depicted  viable parameter space  for the model parameters $\alpha$ and $\beta$.  From \fig{fig_rns-pactlb} and \fig{fig_betamax_pact_LBD}, we conclude that the joint constraint P-ACT-LB+BK18, when combined with the LitBIRD sensitivity, favours the EOS parameter space with $1 \simeq \alpha < 2$ i.e. models between monomial large-field inflation and Starobinsky/$\alpha$-attractor, wherein $\beta$ remains $\mathcal{O}(1)$ for larger values of $\alpha$, while shifting to $\mathcal{O}(10^{-1})$ for smaller $\alpha$. Here also plateau-type (hilltop) inflationary scenario remains outside the observationally viable region.

\begin{figure}
	\centering
	\begin{subfigure}{.5\textwidth}
		\centering
		\includegraphics[width=.97\linewidth, height=7cm]{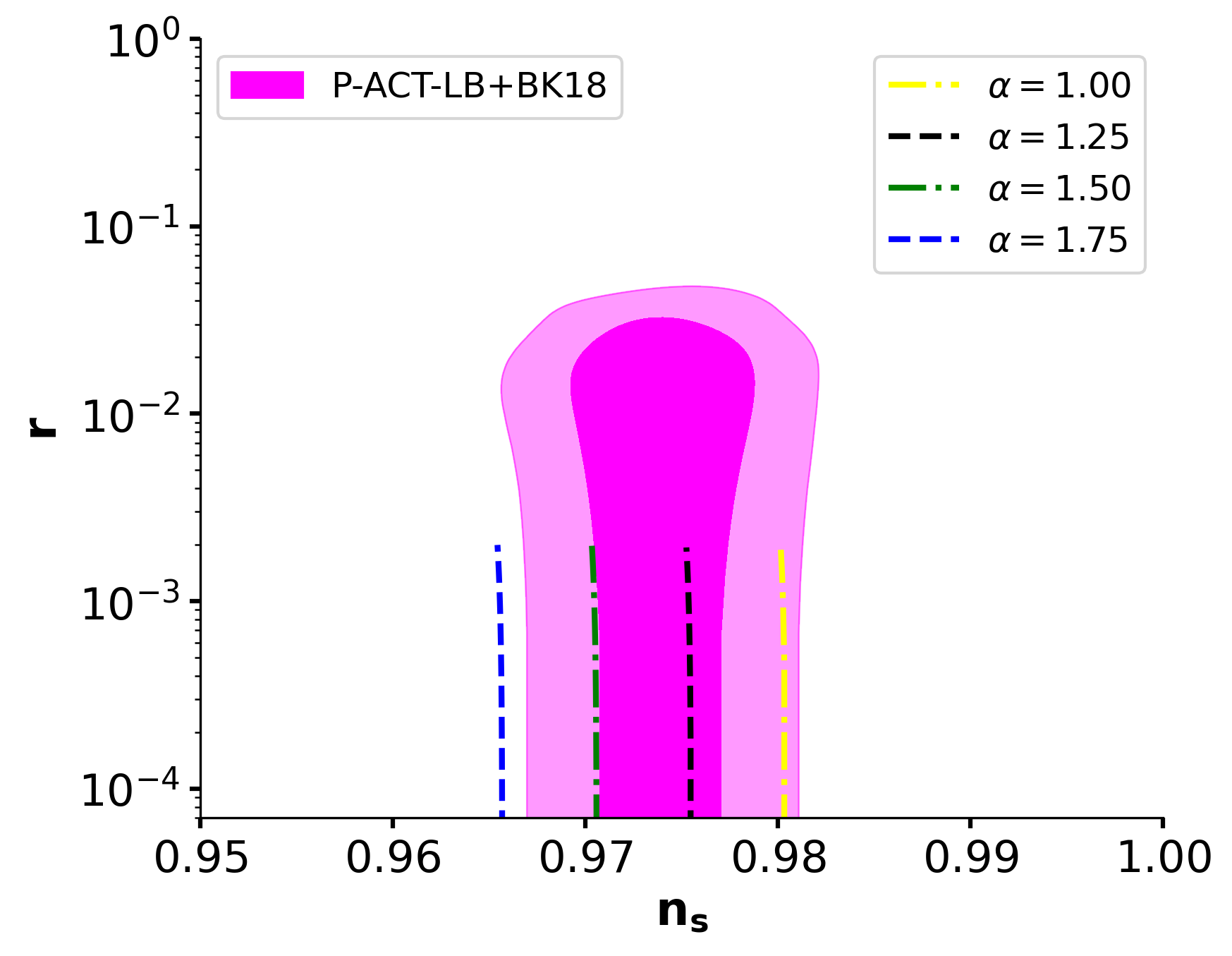}
		\caption{N=50: LiteBIRD No-Detection}
	\end{subfigure}%
	\begin{subfigure}{.5\textwidth}
		\centering
		\includegraphics[width=.97\linewidth, height=7cm]{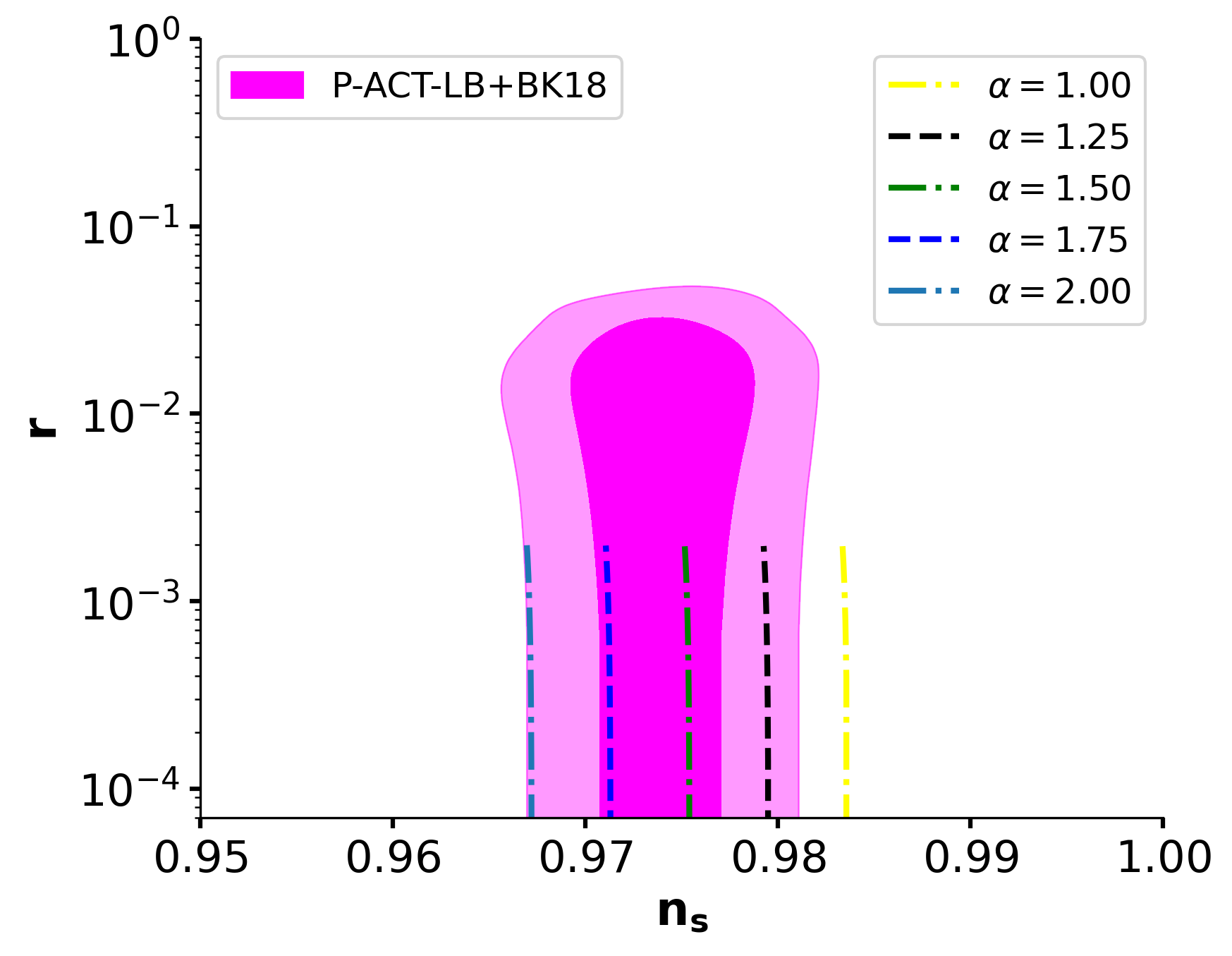}
		\caption{N=60: LiteBIRD No-Detection}
	\end{subfigure}
	\caption{The $68\%$ and $95\%$ confidence regions in the $r$--$n_{_S}$ plane obtained from the joint analysis of P-ACT-LB+BK18. The dashed curves denote the EOS inflation predictions for different fixed values of the model parameter $\alpha$, two values of number of e-folding and varying the other model parameter $\beta$ within its allowed range determined by $r<0.002$. }
	\label{fig_rns-pactnlb}
\end{figure}

\begin{figure}
	\centerline{\includegraphics[width=15.cm, height=8cm]{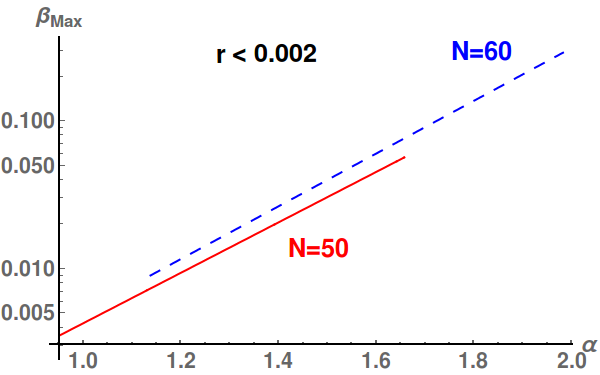}}
	\caption{\label{fig_betamax_pactnlb}Variation of $\beta_{\rm Max}$ with the model parameter $\alpha$ for $N=50$ (solid red line) and $N=60$ (dashed blue curve). For the plot we have considered the constraint on scalar spectral index from the combined analysis of P-ACT-LB data along with  $r<0.002$ from LiteBIRD predictions for non-detection of primordial gravitational waves.}
\end{figure}

The non-detection of primordial gravitational waves by LiteBIRD will push back the upper-limit of tensor-to-scalar ratio to $r<0.002$. In \fig{fig_rns-pactnlb}, we present the EOS inflation predictions for multiple fixed values of $\alpha$ and for two representative choices of the number of e-folds. The parameter $\beta$ is varied across its observationally allowed range, as constrained by the upper bound $r < 0.002$. We have also depicted the marginalized joint confidence contours of  $68\%$ and $95\%$  in $r-n_{_S}$ plane from the combined analysis of P-ACT-LB. In \fig{fig_betamax_pactnlb}, we have plotted the variation of $\beta_{\rm Max}$ with the other model parameter $\alpha$, imposing the restriction $r < 0.002$ alongside the 2-$\sigma$ constraint on $n_{_S}$ from P-ACT-LB. The figure depicts that window for one of the model parameter, $\beta$, has been narrowed owing to fact that non-detection of primordial gravity waves by LiteBIRD will set $r<0.002$. Thus, accommodating LiteBIRD’s anticipated non-detection of primordial gravitational waves necessitates an exceptionally small value of $\beta$, with the remaining parameter exhibiting no significant change.

\begin{figure}
	\centering
	\begin{subfigure}{.5\textwidth}
		\centering
		\includegraphics[width=.97\linewidth, height=7cm]{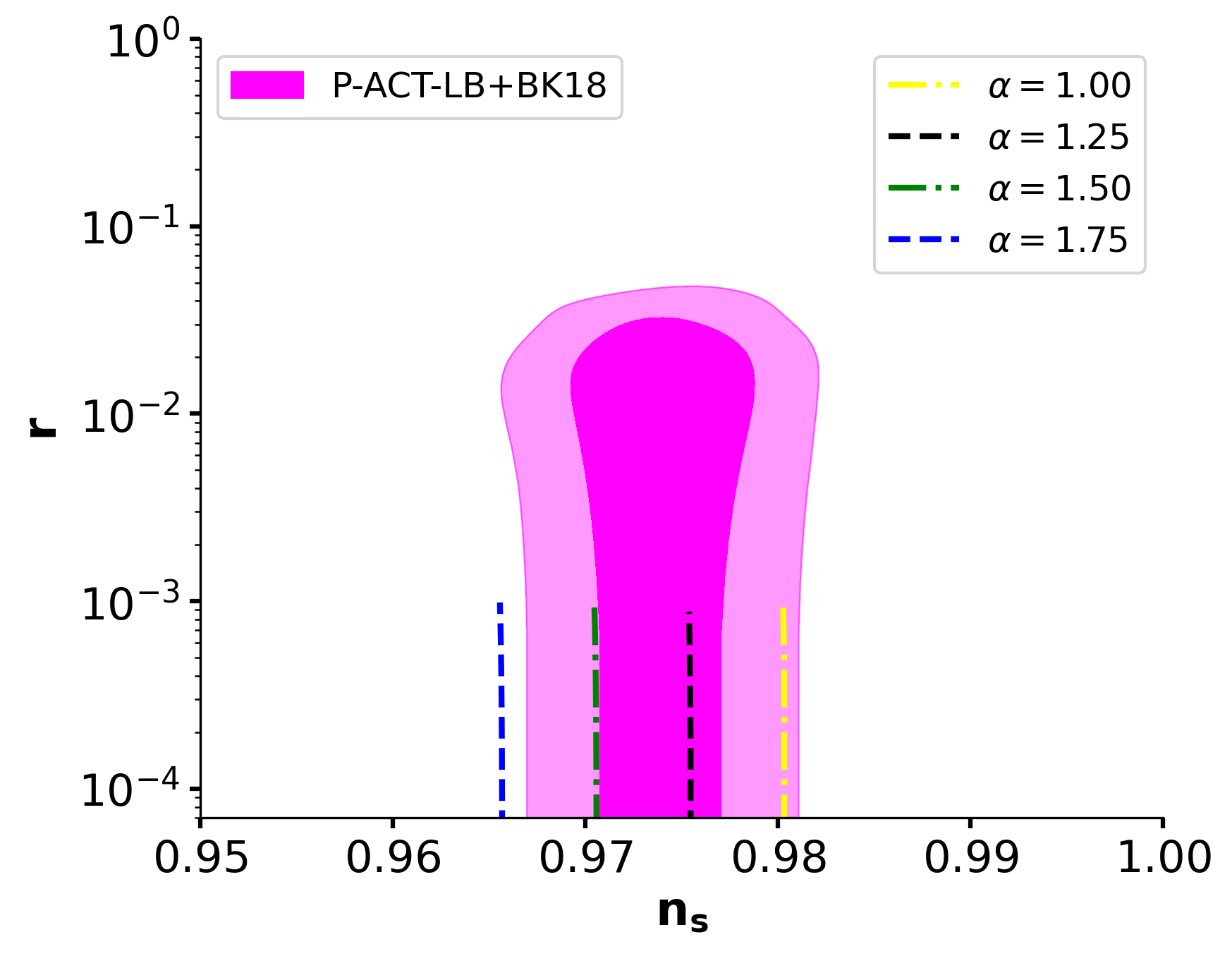}
		\caption{N=50: CMB-S4 No-Detection}
	\end{subfigure}%
	\begin{subfigure}{.5\textwidth}
		\centering
		\includegraphics[width=.97\linewidth, height=7cm]{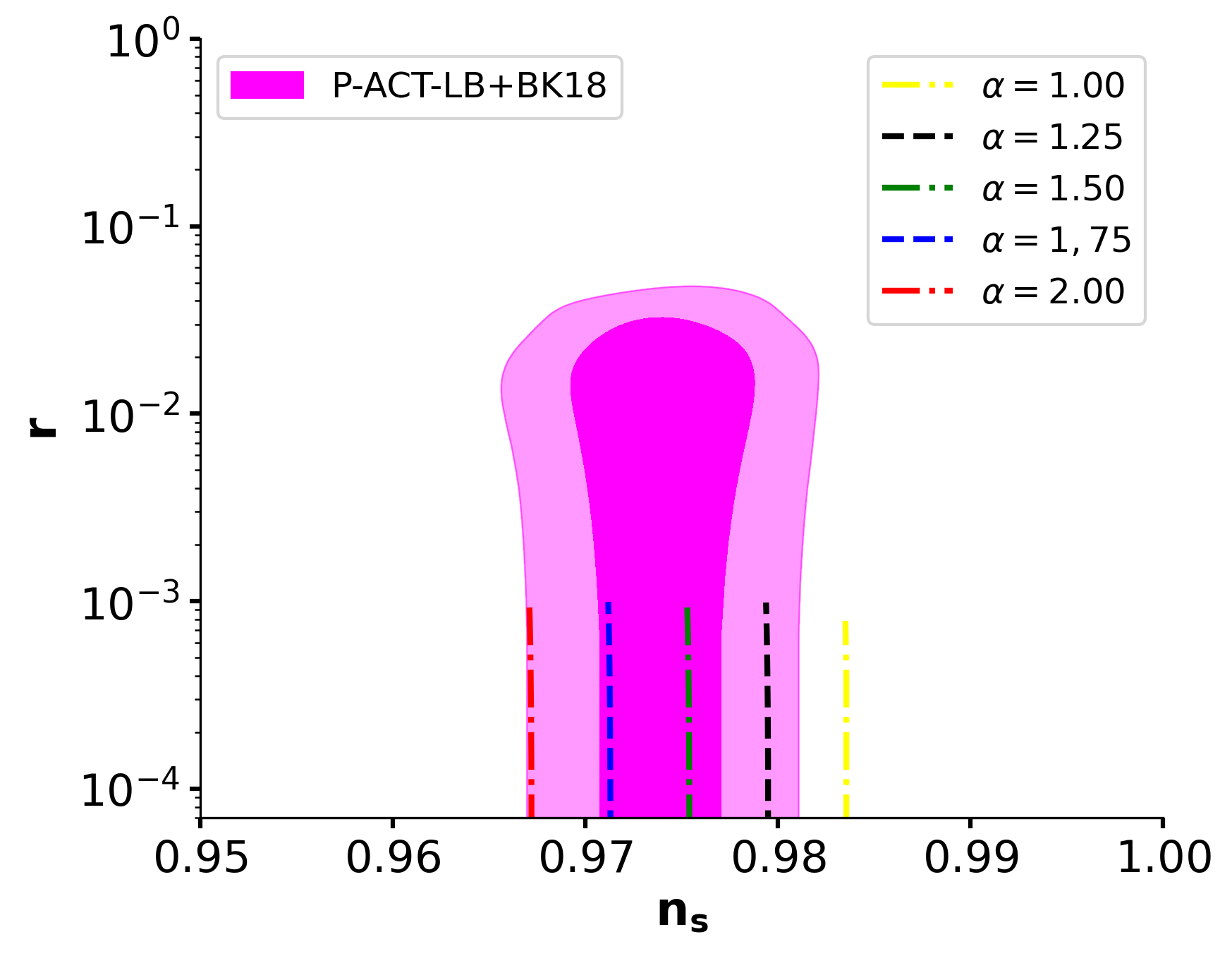}
		\caption{N=60: CMB-S4 No-Detection}
	\end{subfigure}
	\caption{The $68\%$ and $95\%$ confidence regions in the $r$--$n_{_S}$ plane obtained from the joint analysis P-ACT-LB+BK18. The dashed curves denote the EOS inflation predictions for different fixed values of the model parameter $\alpha$, two different  number of e-folding and varying the other model parameter $\beta$ within its allowed range determined by $r<0.001$. }
	\label{fig_rns-pactns4}
\end{figure}

\begin{figure}
	\centerline{\includegraphics[width=15.cm, height=8cm]{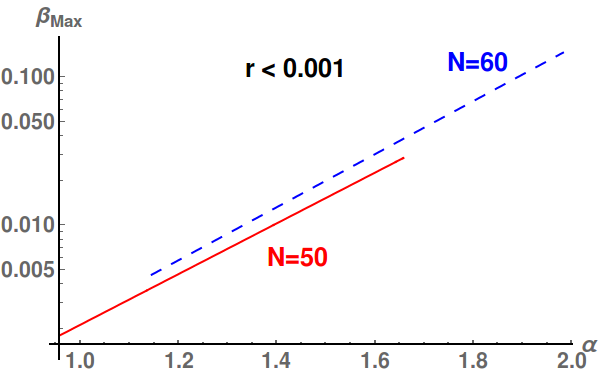}}
	\caption{\label{fig_betamax-pactns4}Variation of $\beta_{\rm Max}$ as a function of $\alpha$ for $N = 50$ and $N = 60$. The constraints used in the plot include the scalar spectral index bounds from the combined P–ACT–LB data and the requirement $r < 0.001$, consistent with CMB-S4 forecasts assuming a non-detection of primordial gravitational waves.}
\end{figure}

CMB-S4  is a ground based mission expected to detect tensor perturbation within the range $0.003<r<0.032$ \cite{abazajian2016cmb, matsumura2014mission,abazajian2022cmbS4,belkner2024cmbs4}. However, failure of CMB-S4 mission to observe primordial gravity waves would imply $r<0.001$. Since the range for detection of gravity waves by LiteBIRD and CMB-S4 are identical, they would provide similar constraints on the model parameter and hence we do not repeat  our analysis for CMB-S4. But in case of non-detection of primordial gravity waves by CMB-S4, the observationally viable window for the model parameter $\beta$ is further shrunk while allowed range of  $\alpha$ remaining almost invariant as depicted in \fig{fig_rns-pactns4} and \fig{fig_betamax-pactns4}.

Therefore, the non-detection of primordial gravitational waves by LiteBIRD and/or CMB-S4 primarily tightens the parameter space of $\beta$, restricting it from the broad range previously allowed to a much narrower window of $\mathcal{O}(10^{-3})$ to $\mathcal{O}(10^{-1})$. In contrast, the viable range of $\alpha$ remains largely unchanged compared to earlier analyses. 


\section{Conclusion}
In this article, we have explored the inflationary equation-of-state framework in the context of the latest ACT-DR6 observations. We find that the Mukhanov parametrization of inflationary equation-of-state does provide an excellent match with recent ACT-DR6 data. We also find that  one of the model parameter, $\beta$, is highly sensitive to the  tensor-to-scalar ratio along with the number of e-foldings.  While the range for $\alpha$ is mainly dependent on the constraint for scalar spectral index and $N$. Moreover,  observationally viable  range of the model parameter $\beta$ is reduced considerably when we also utilize the prediction for non-detection of primordial gravity waves from the futuristic CMB experiments  LiteBIRD and/or CMB-S4.  The parameter space for equation-of-state formalism is significantly reduced when non-detection of primordial gravity waves from the futuristic CMB experiments  LiteBIRD and/or CMB-S4 is also taken into consideration. 

We find that the region $\alpha\lesssim 1$, corresponding to power-law or monomial inflation, which Planck has strongly disfavoured, re-enters the allowed parameter space, though for very small $\beta$, when the full P–ACT–LB+BK18 joint constraints are taken into account. Moreover, the EOS model remains viable regardless of whether future CMB experiments such as LiteBIRD and/or CMB-S4 detect or fail to detect primordial gravitational waves. However, consistency with these future sensitivities requires the parameter $\beta$ to lie in a very small range, typically  $\sim\mathcal{O}(10^{-2})$.
While, the range  $1<\alpha<2$ corresponds to inflationary models that interpolate between power-law/monomial behaviour and plateau-type scenarios remains fully consistent with P-ACT-LB+BK18, for low to moderately large $\beta$ values, even for non-detection of primordial gravitational waves by LiteBIRD and/or  CMB-S4. By contrast, the Starobinsky model (\(\alpha = 2\)) sits near the boundary of the $95\%$ contour, with its compatibility sensitive to the choice of the e-folding number. We additionally observe that hilltop/small-field inflation models ($\alpha>2$  are pushed outside the $95\%$ confidence contour of the P–ACT–LB–BK18 combination.

In a nut shell, we find the  equation-of-state formalism provides adequate parameter space where the model  can imitate  recent data from various observational probes in the likes of Planck-2018, DESI-Y1, ACT-DR6 along with their combinations. However the non-detection of primordial tensor perturbations by LiteBIRD and/or CMB-S4 will significantly reduce the range for $\beta$ while keeping the observational window for $\alpha$ almost fixed.

\section*{Acknowledgment}
I would like to sincerely thank anonymous reviewers for their critical and constructive suggestions on the first version of this work.  Which help me a lot to improve the quality of this paper. 


\begingroup
\bibliographystyle{unsrt}
\bibliography{references}
\endgroup

\end{document}